\documentclass[11pt]{article}
\usepackage{jheppub}
\usepackage{tensor}
\usepackage{amsmath,amssymb,amsfonts,graphicx,tensor}
\usepackage{times,palatino,newpxmath,courier}
\usepackage{subcaption}

\newcommand{\dd}{\mathrm{d}}
\usepackage{graphicx}
\usepackage{enumitem}
\usepackage{braket}
\usepackage[dvipsnames]{xcolor}
\usepackage{comment}

\title{Bubbles of cosmology in AdS/CFT}
\author{Abhisek Sahu, Petar Simidzija, Mark Van Raamsdonk}
\affiliation{Department of Physics and Astronomy, University of British Columbia,\\
6224 Agricultural Road, Vancouver, B.C., V6T 1Z1, Canada}

\emailAdd{abhi@phas.ubc.ca}
\emailAdd{psimidzija@phas.ubc.ca}
\emailAdd{mav@phas.ubc.ca}
\date{March 2022}

\abstract{Gravitational effective theories associated with holographic CFTs have cosmological solutions, which are typically big-bang / big-crunch cosmologies. These solutions are not asymptotically AdS, so they are not dual to finite-energy states of the CFT. However, we can find solutions with arbitrarily large spherical bubbles of such cosmologies embedded in asymptotically AdS spacetimes where the exterior of the bubble is Schwarzschild-AdS. In this paper, we explore such solutions and their possible CFT dual descriptions. Starting with a cosmological solution with $\Lambda < 0$ plus arbitrary matter density, radiation density, and spatial curvature, we show that a comoving bubble of arbitrary size can be embedded in a geometry with AdS-Schwarzschild exterior across a thin-shell domain wall comprised of pressureless matter. We show that in most cases (in particular, for arbitrarily large bubbles with an arbitrarily small negative spatial curvature) the entropy of the black hole exceeds the (radiation) entropy in the cosmological bubble, suggesting that a faithful CFT description is possible. We show that unlike the case of a de Sitter bubble, the Euclidean continuation of these cosmological solutions is sensible and suggests a specific construction of CFT states dual to the cosmological solutions via Euclidean path integral.
}

\begin{document}

\maketitle

\section{Introduction}

The holographic approach to quantum gravity provides a fully microscopic definition for certain quantum gravitational theories in terms of dual non-gravitational quantum systems. The best understood examples describe gravitational physics in spacetimes that are asymptotically empty and negatively curved. On the other hand, observations suggest that our own universe is qualitatively different, an approximately homogeneous and isotropic expanding spacetime filled with matter and radiation. It is a challenge to understand whether the holographic approach can be extended to describe such cosmological spacetimes.\footnote{Various approaches have been considered in the past. See \cite{Banks:2001px,Strominger:2001pn,Alishahiha:2004md,Gorbenko:2018oov,Coleman:2021nor,Freivogel2005,McFadden:2009fg,Banerjee:2018qey,Susskind:2021dfc, Hertog:2005hu} for examples.}
\begin{figure}
    \centering
    \begin{subfigure}{0.3\textwidth}
    \includegraphics[scale=0.5]{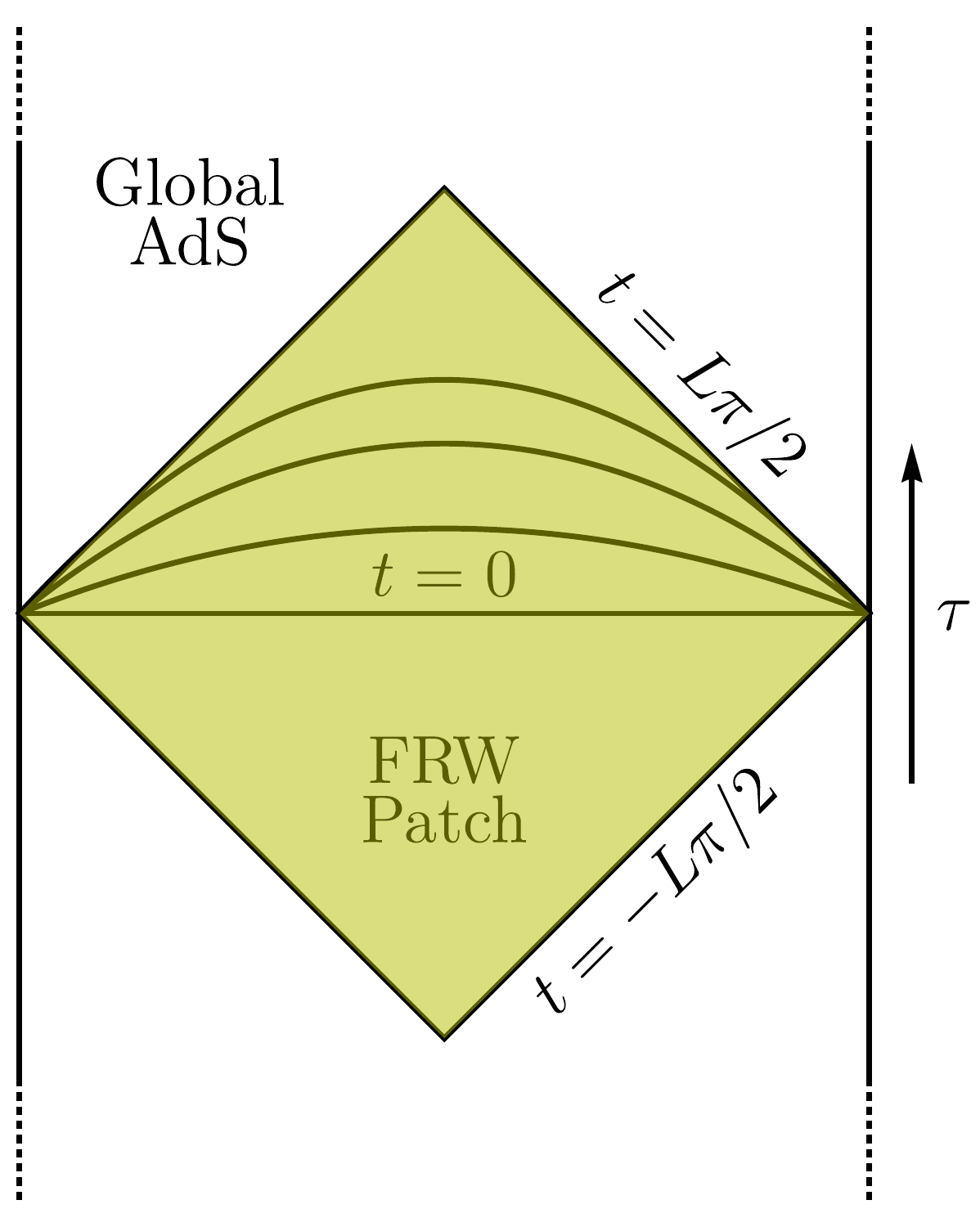}
    \end{subfigure}
     \hspace{0.2\textwidth}
    \begin{subfigure}{0.3\textwidth}
     \centering
     \includegraphics[scale=0.5]{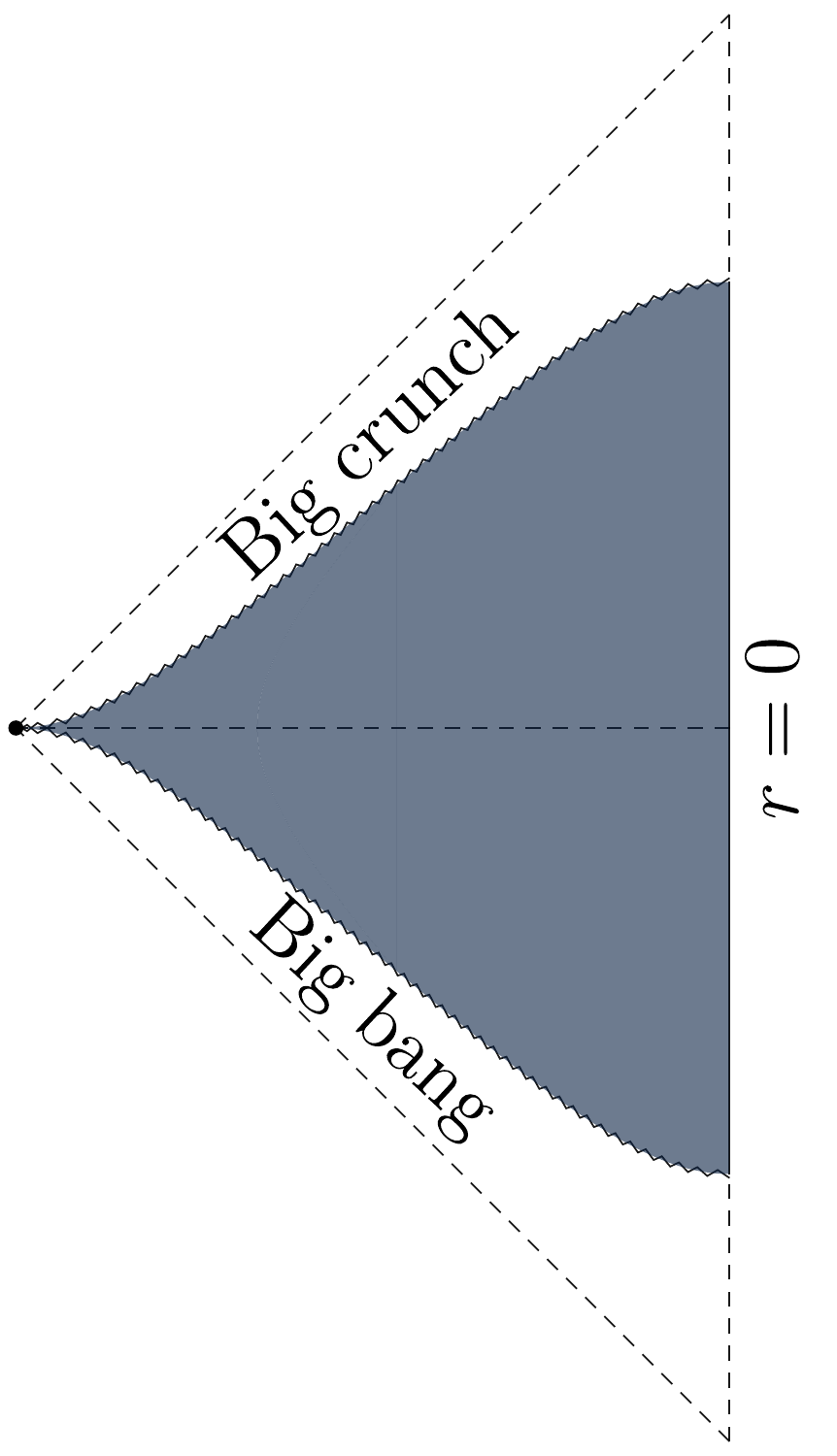}
     \end{subfigure}
    \caption{Left: FRW patch of Global AdS spacetime showing constant FRW time Cauchy slices. The surfaces $t = \pm L \pi/2$ represent coordinate singularities. Right: Adding an arbitrarily small matter or and/or radiation density gives a $\Lambda < 0$ big-bang big crunch cosmology with Penrose diagram shown here. The spacetime is no longer asymptotically AdS.}
    \label{fig:AdSPatch}
\end{figure}

In this paper, we will consider $\Lambda < 0$ cosmological spacetimes that are most similar to spacetimes that we can already describe holographically. Pure AdS itself is an example of a spacetime that can be written in FRW form (with negative spatial slices) as
\begin{equation}
 -dt^2 + L^2 \cos^2(t/L) dH^2_{n-1}
\end{equation}
where $dH^2_{n-1}$ is the unit hyperbolic metric. This FRW metric covers only a patch of the full global AdS spacetime (see figure \ref{fig:AdSPatch}), equivalent to the domain of dependence of a constant global-time slice of global AdS. The dual CFT vacuum state provides a holographic description of this trivial cosmology. The ``big bang'' and ``big crunch'' at $t = \mp L \pi / 2$ are only coordinate singularities in this case. However, by adding an arbitrarily small homogeneous and isotropic density of matter and/or radiation to the $t=0$ slice, we obtain a big bang/big crunch cosmology with genuine curvature singularities. The spacetime diagram for the resulting cosmology is shown in Figure \ref{fig:AdSPatch}. This spacetime is no longer asymptotically AdS, so the new spacetime does not correspond to a state in the dual CFT. This is understandable: while the energy {\it density} that we are adding to the $t=0$ slice can be arbitrarily small, the total energy on the infinite volume slice is infinite, so would correspond to a state of infinite energy from the CFT perspective.

\begin{figure}
\centering
\includegraphics[width=0.4\textwidth]{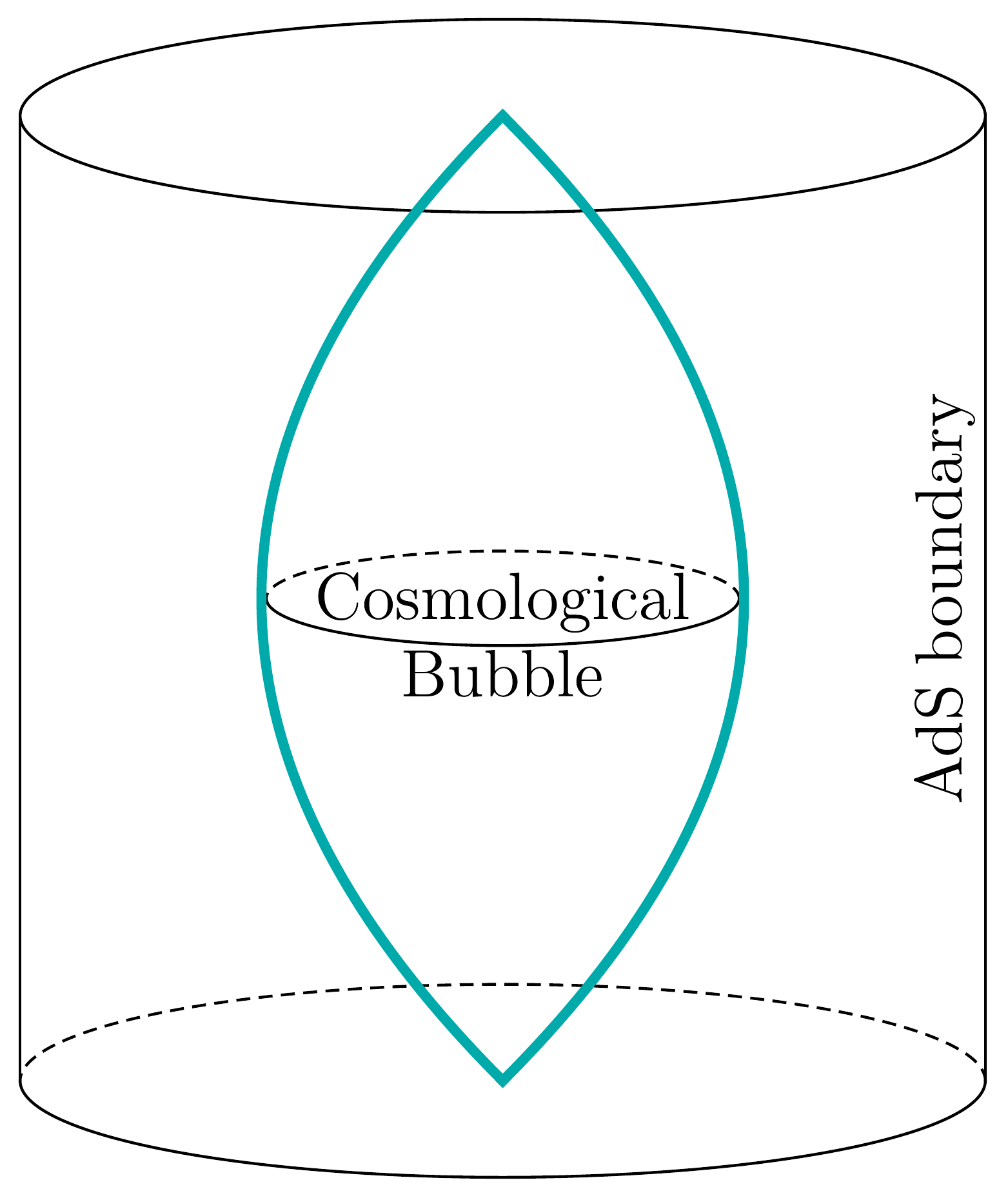}
\hspace{0.05\textwidth}
\includegraphics[width=0.4\textwidth]{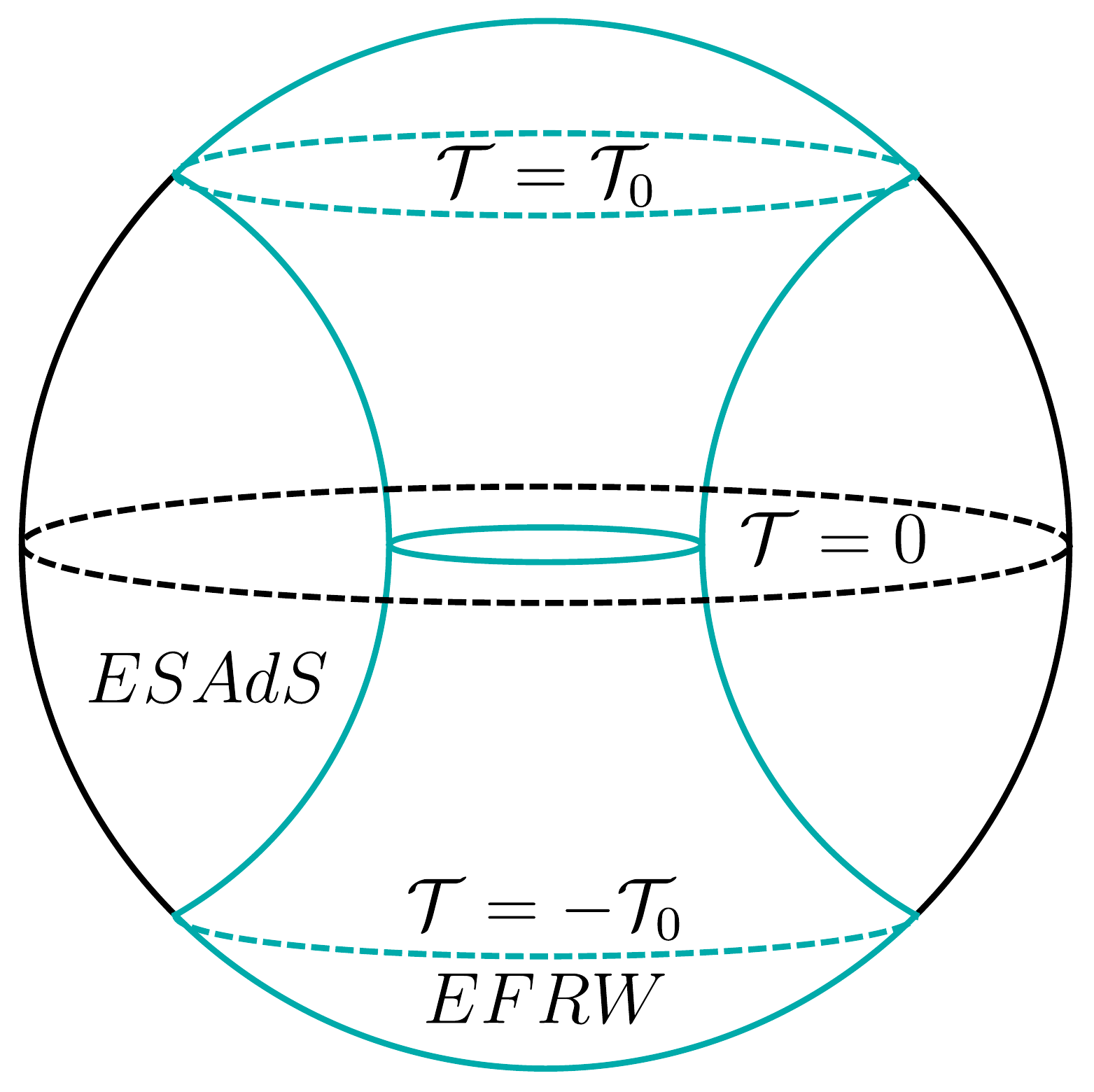}
\caption{Left: schematic representation of the Lorentzian spacetime where the interior geometry is FRW and the exterior geometry is a black hole. Right: Euclidean continuation where the interior region is part of an AdS Euclidean wormhole.}
\label{fig:Bubble_Schematic}
\end{figure}

In order to make progress, we could attempt to define the cosmology from the CFT by taking a limit. Instead of including a uniform density of matter and/or radiation everywhere on the $t=0$ slice, we could have a finite-sized ball filled with a uniform density of matter and/or radiation, with empty negatively curved spacetime outside. The full spacetime obtained by taking this slice as initial data will contain an interior region that is a patch of the FRW cosmology and an exterior region that is a part of a Schwarzschild AdS of some mass (Figure \ref{fig:Bubble_Schematic}). In between these two regions will be a ``domain wall'' that interpolates between the two spacetimes.\footnote{This doesn't have to be a physical object; we are using the term to indicate a transition region in the geometry that is neither FRW or Schwarzchild-AdS.} If the ball is large enough, the interior region will include the the full causal patch associated to a complete geodesic from the big bang to the big crunch\footnote{We could also choose to include a larger patch that contains the union of the causal past and the causal future of the observer rather than just the intersection.}, so one might argue that a holographic description of such a spacetime would be entirely satisfactory for describing the physics accessible to any single observer in the cosmology.

Since these bubble-of-cosmology spacetimes are again asymptotically AdS, it is plausible that they correspond to valid finite-energy states of the dual CFT. These states can be understood as special microstates of the bulk black hole geometry where the black hole interior includes a patch of cosmological spacetime. 

The goal of this paper is to construct such solutions for various types of cosmologies, to investigate whether these solutions can arise from legitimate dual CFT states, and to ask in this case how the cosmological physics is encoded in the dual CFT. Related earlier work along these lines is described below.

We now present a brief outline and summary of the results. 
\paragraph{Constructing solutions}
In section 2, we construct a family of cosmological bubble solutions. The interior geometry is a comoving bubble (i.e. with fixed coordinate radius in FRW coordinates) of a cosmological solution with arbitrary spatial curvature, matter density, and radiation density. The exterior geometry is a part of the maximally extended AdS-Schwarzschild solution. We find that (in four dimensions) these two regions can always be joined across a domain wall that is a thin shell of dust whose density depends only on the radiation density in the interior.

Depending on the parameters, the cosmological bubble may lie completely behind the horizon of the black hole, or may start outside the horizon and fall through the horizon at some later time.

\paragraph{Potential entropy puzzle}

For sufficiently small bubbles, it seems clear that there should be a CFT description - the spacetimes just correspond to a small amount of matter and/or radiation localized in some region of an otherwise empty AdS. However, for larger bubbles, there is a potential puzzle. Since the exterior spacetime is Schwarzschild-AdS, it has a naturally associated Bekenstein-Hawking entropy given by $1/(4G)$ times the area of the horizon. We can also associate an entropy to the interior of the bubble. For example, in a universe with radiation, this interior entropy should be bounded below by the entropy of the radiation. As we increase the size of the bubble, the interior entropy might eventually exceed the black hole entropy (plausible since the interior entropy scales as a volume but the exterior entropy scales like an area). In this case, the CFT Hilbert space (restricted to the relevant energy band) would not be large enough to accommodate all the interior radiation microstates as independent states. Then either the bulk geometries do not correspond to valid CFT states, or the black hole interiors must be encoded non-isometrically (as in \cite{Akers:2022qdl}). 

We investigate these entropies in section 3. We find that with arbitrarily small negative spatial curvature, the black hole entropy always exceeds the radiation entropy in the cosmology. This is natural since for negative curvature, volume and area are proportional. For flat cosmologies, the radiation entropy eventually exceeds the black hole entropy, but only for bubbles that are parametrically larger than the cosmological scale (by a factor that is a positive power of $\ell_P/\ell_{AdS}$). There is never an issue with describing the region accessible to a single observer.   

\paragraph{Euclidean construction}

Another hint about the existence or not of a dual CFT state for a cosmological bubble geoemetry comes by looking at whether the solution has an analytic continuation to a well-behaved Euclidean geometry. If it is well-behaved, the corresponding Euclidean solution might represent the saddle point of a gravitational path integral whose dual CFT description constructs the CFT state. For de Sitter bubbles embedded in AdS (considered originally in \cite{freivogel_inflation_2006}), the analytic continuations are not sensible since they have self-intersecting domain walls \cite{fu_bag--gold_2019}. In our case, we find in section 4 that the domain wall trajectories are sensible in the Euclidean picture for all cases with $K \le 0$ and for any bubble smaller than half of the full spatial sphere in the $K > 0$ case.

When the Euclidean solution is sensible, the boundary has the topology of a sphere $S^3$, and the Euclidean domain wall has the topology of a cylinder ($S^2 \times I$) that intersects the sphere on a pair of $S^2$s, as shown in Figure \ref{fig:Bubble_Schematic}. As we discuss in section 5, this suggests that the Euclidean geometry arises holographically from a Euclidean CFT on $S^3$ with insertions of sources and/or operators restricted to a pair of polar regions. As a simple example, we can consider the case with pure matter, where the matter is a collection of heavy particles sitting at fixed coordinate locations in the FRW description. In this case, the CFT description involves the insertion of heavy CFT operators as specific locations on the sphere corresponding to the endpoints of the geodesics in the Euclidean picture, similar to the construction of shells of matter in \cite{,Anous:2016kss,Balasubramanian:2022gmo}. In order that the desired Euclidean solution provides the dominant saddle of the CFT path integral with such insertions, it may be necessary to consider an ensemble of such insertions with a certain degree of correlation between the insertions on the two sides of the sphere. We discuss this in more detail in the section 5.

\paragraph{RT surface probes}

If a bubble geometry does have a valid CFT state, it is interesting to ask how we can probe the cosmological region inside the bubble using CFT quantities. The spherical boundary at a given time is contractible in the bulk geometry, so the entanglement wedge of the boundary is the entire spacetime. This suggests that reconstruction is possible in principle. However, in the cases where the cosmology is entirely behind the black hole horizon, the horizon provides a non-minimal extremal surface for the full boundary. In this case, the cosmology is in a ``Python's lunch,'' and the general arguments of \cite{Brown:2019rox} suggest that accessing the cosmology from the CFT state will be exponentially complex. We give a precise condition for when this occurs (equation 2.19). We also investigate the behavior of RT surfaces for ball-shaped regions in the CFT to gain insight into how much of the cosmological region is encoded in proper subsystems of the boundary. 

Even if the cosmology is fully behind the horizon in a Python's lunch, if there is a Euclidean construction of the state as described above, we should be able to use this Euclidean picture in order to calculate cosmological observables in a straightforward way. Thus, extracting the physics from the underlying CFT physics can be simple even if the extraction from the Lorentzian CFT state is complex.

\paragraph{Relation to earlier work}

Describing cosmology as a bubble in AdS was considered previously in \cite{freivogel_inflation_2006}, where the authors described solutions with a bubble of de Sitter spacetime embedded in AdS with thin constant tension bubble wall. Though there is no problem in constructing the solutions, it is unclear whether these geometries are dual to legitimate CFT states. In particular, \cite{fu_bag--gold_2019} found evidence that these solutions cannot arise from a Euclidean path integral construction of states since the domain walls appearing in the Lorentzian geometries are badly-behaved in the Euclidean continuation.

In \cite{fu_bag--gold_2019,simidzija_holo-ween_2020}, the authors considered bubbles of vacuum AdS in exterior asymptotically AdS spacetimes with a different cosmological constant. In this case, the solutions often do have sensible Euclidean continuations, suggesting a Euclidean path integral construction of the states that involves an interface between CFTs. These can be viewed as constructions of bubbles of a trivial cosmology. 

With other collaborators, some of us have considered a different approach to holographically constructing $\Lambda <0$ cosmologies in \cite{Cooper2018,VanRaamsdonk:2021qgv,Antonini2022short}. We comment in the discussion section about the relation between those works and the construction of the present paper.

\section{Spherical Bubbles of Cosmology}

In this section, we construct a family of solutions with an interior region that is a bubble of FRW $\Lambda < 0$ radiation + matter big-bang / big crunch cosmology with a comoving boundary, an exterior region that is AdS/Schwarzchild black hole, and a boundary shell of dust / non-relativistic matter. We will assume that the cosmological constant is the same throughout the spacetime, taking $\Lambda = - 3/\ell_{AdS}^2$. We will use units where $\ell_{AdS} = 1$. 

Inside the bubble, we have an FRW metric
\begin{equation}
    ds^2 = -dt^2 + a^2(t) \left(dr^2 + R_K(r)^2 (d\theta^2+\sin^2{\theta}d\phi^2)\right) \quad R_K(r) = \left\{ \begin{array}{ll} \sin(\sqrt{K} r)/\sqrt{K} & K > 0 \cr r & K=0 \cr \sinh(\sqrt{|K|} r)/\sqrt{|K|} & K < 0 \end{array} \right.
\end{equation}
where the scale factor $a(t)$ is taken to be dimensionless and $K$ is the Gaussian curvature of the spatial slices. The scale factor satisfies the Friedmann equation
\begin{equation}
\label{Friedmann}
\left({\frac{\dot{a}}{a} }\right)^2  =  {\frac{\rho_M}{a^3} } + {\frac{\rho_R}{a^4}} -{\frac{K} {a^2}}  - 1 \; ,
\end{equation}
where a dot denotes a $t$ derivative. As the universe expands, the matter and radiation dilute so eventually there is a time where $\dot{a}=0$, after which the universe contracts to a big crunch. The full solutions are time-reversal symmetric about this recollapse point. Without loss of generality, we will define this time to be $t=0$ and choose $a(0) = 1$. In this case, $\rho_R$ and $\rho_M$ represent the radiation and matter densities at $t=0$ as a fraction/multiple of the ``critical'' energy density  $3/(8 \pi G \ell_{AdS}^2)$ for a flat cosmology. The Friedmann equation at $t=0$ gives
\begin{equation}
\label{Krho}
 K = \rho_M  + \rho_R  - 1
\end{equation}
so the spatial curvature at the time-symmetric slice is fixed in terms of the matter and radiation densities. For empty AdS, we get $K=-1$ as expected.

\begin{figure}
    \centering
\includegraphics[scale=0.5]{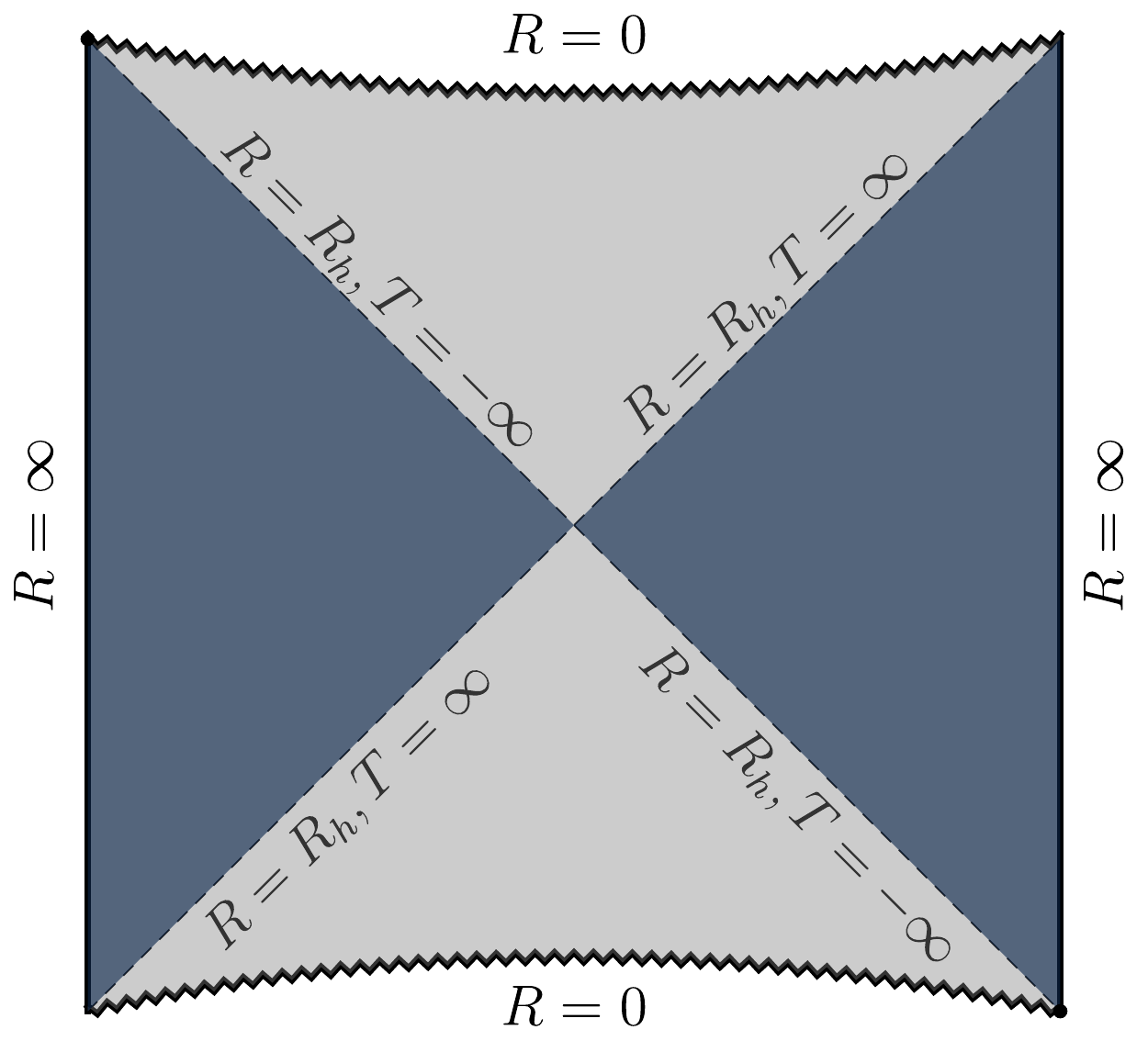}
    \caption{Penrose diagram of a maximally extended AdS Schwarzschild black hole. }
    \label{fig:SAdSPenrose}
\end{figure}

We assume that the space outside of the ball is empty and spherically symmetric, so by Birkhoff's theorem, the domain of dependence of the exterior of the ball must be a part of the maximally extended Schwarzschild-AdS geometry. We can use Schwarzchild coordinates 
\begin{align}
    dS^2 &= -F(R) dT^2 +\frac{dR^2}{F(R)} + R^2 (d\theta^2+\sin^2{\theta}d\phi^2) \; ,
\end{align} 
to describe the various patches of that geometry, where 
\begin{equation}
\label{defF}
F(R) = R^2 + 1 - {\frac{\mu}{R}} \; .
\end{equation}
Here, is related to the mass of the black hole by  $\mu = 2 M G_N$. Defining the horizon radius by $F(R_h)=0$, we have that the past and future wedges are described using coordinates in the range $R\in(0,R_h), T\in (-\infty,\infty)$ (where $R=0$ represents the past/future singularity), while the left and right wedges are described using coordinates in the range $R\in(R_h,\infty), T\in (-\infty,\infty)$ (Figure \ref{fig:SAdSPenrose}).

In our ansatz, the surface of the bubble is assumed to be comoving with respect to the FRW interior, so it has fixed coordinate radius $r(t) = r_0$. It is natural to use the FRW coordinates $(t,\theta,\phi)$ to parameterize the boundary surface in the FRW spacetime.  The intrinsic metric on the bubble is then
\begin{equation}
\label{induced}
 ds^2 = -dt^2 + a^2(t) R_K(r_0)^2 d \Omega_2^2 \; .   
\end{equation}
We will consider a shell of dust / non-relativistic matter living on the bubble. This can be described via the stress-energy tensor
\[
8 \pi G_N S^t {}_t = - {\rho_s \over a(t)^2} \; .
\]
Note that the energy density dilutes as $1/a^2$ since the bubble is 2+1 dimensional.

We require that the Israel junction conditions be satisfied for the bubble. The first junction condition requires that the induced metric from the exterior spacetime agrees with the induced metric (\ref{induced}) from the interior spacetime. Thus, the trajectory of the bubble in the exterior spacetime can be described as $(R(t), T(t))$ where we require that
\begin{equation}
\label{eq:SchBubble}
R(t) =  a(t) R_K(r_0) \qquad \qquad F(R(t))\dot{T}^2 - {\dot{R}^2 \over F(R(t))} = 1
\end{equation}
The second junction condition requires that the difference of extrinsic curvatures between the interior and exterior spacetimes is determined by the surface stress-energy tensor as
\begin{equation}
\label{SecondJC}
    K_{ab}^{in} - K_{ab}^{out} = 8 \pi G_N(S_{ab} - {1 \over 2} h_{ab} S)
\end{equation}
where $K_{ab} = \nabla_\mu n_\nu e_a^\mu e_b^\nu$, and $n$ is a unit normal vector to the surface in the relevant spacetime, chosen to point outward (away from the FRW interior) in both cases.
In the FRW geometry, we can describe an outward pointing unit normal vector to the surface as 
\[
n_\mu = (0, a(t) ,0,0) \; .
\]
From this, we find that the extrinsic curvature has nonzero components
\begin{eqnarray}
    K^{\theta} {}_{\theta} &=& - h^{\theta \theta} \Gamma^r_{\theta \theta} n_r = {1 \over 2} g^{rr} \partial_r g_{\theta \theta} n_r =  {R_K'(r_0) \over R_K(r) a(t)} \\
    K^{\phi} {}_{\phi} &=& - h^{\phi \phi} \Gamma^r_{\phi \phi} n_r = {1 \over 2} g^{rr} \partial_r g_{\phi \phi} n_r = {R_K'(r_0) \over R_K(r) a(t)} \\
\end{eqnarray}
In the Schwarzschild geometry near the $t=0$ slice, the outward pointing unit normal vector to the surface is 
\[
n_\mu = \epsilon(-\dot{R},\dot{T},0,0) \; .
\]
where $\epsilon = 1$ if the $n$ is outside the Schwarzschild horizon pointing towards the asymptotic region, and $\epsilon = -1$  if $n$ is inside the Schwarzchild horizon pointing outward (i.e. if the bubble is completely inside the horizon). In this case, we can write the nonzero components of the extrinsic curvature as 
\begin{equation}
K^\theta {}_\theta = K^\phi {}_\phi = {\beta \over R_K(r_0) a(t)} \qquad
K^t {}_t  = {\dot{\beta} \over R_K(r_0) \dot{a}(t)} \qquad
\beta \equiv  \epsilon \sqrt{F[a(t) R_K(r_0)] + R_K(r_0)^2 \dot{a}^2(t)} \; .
\end{equation}
We are now ready to write the second junction condition. The ${}^\phi {}_\phi$ and ${}^\theta {}_\theta$ components of (\ref{SecondJC}) give
\begin{equation}
\label{junction}
{R_K'(r_0) \over R_K(r_0) a(t)} - {\beta \over R_K(r_0) a(t)} = {1 \over 2} {\rho_S \over a^2(t)} 
\end{equation}
The ${}^t {}_t$ component gives
\[
- {\dot{\beta} \over R_K(r_0) \dot{a}(t)} = -{1 \over 2} {\rho_S \over a^2(t)}
\]
We can check that this is satisfied provided that (\ref{junction}) is satisfied by solving for $\beta$ and differentiating. 
Thus, we will have a solution provided that 
\begin{equation}
\label{Junction2}
\epsilon \sqrt{F[a(t) R_K(r_0)] + R^2_K(r_0) \dot{a}^2} = R_K'(r_0) - {1 \over 2} {\rho_S R_K(r_0) \over a(t)} \; .
\end{equation}
Squaring this, and substituting the explicit expression for $F$ from (\ref{defF}) and the result for $\dot{a}^2$ from the Friedmann equation  (\ref{Friedmann}), we find the the equation holds only if
\begin{eqnarray}
    \rho_S &=& 2 \sqrt{\rho_R} \cr
    \mu &=& R^3_K(r_0) \rho_M  + 2 \sqrt{\rho_R} R_K^2(r_0) R_K'(r_0)
\end{eqnarray}
where we recall that $r_0$ is the proper distance from the middle of the bubble to the bubble wall at $t=0$ and  $R_0 \equiv R_K(r_0)$ is the proper bubble wall (areal) radius at $t=0$.
The first equation shows that there is a particular density $\rho_S$ of dust on the shell that is required to ``contain'' the cosmology. If there is no radiation, this density goes to zero; the resulting solution (described earlier in \cite{giddings2002gravitational}) is an AdS version of the Oppenheimer-Snyder solution for the collapse of a uniform ball of dust in Minkowski space.

\paragraph{Mass and horizon radius of the black hole}
The second equation above gives the mass parameter of the black hole in terms of the ball radius and the matter and radiation densities inside.\footnote{Using $\mu = 2MG$ and the relationship between $\rho_M$ and $\rho_S$ to the physical energy densities, we see that when $K=0$ this equation says that the black hole mass is precisely equal to the energy of the dust inside the bubble plus the energy of the dust on the bubble. Since dust energy doesn't redshift, this is true at any time slice.} We can also relate the black hole horizon radius to these parameters using the relation
\begin{equation}
\mu = R_H(R_H^2 + 1)  \; .
\end{equation}

We can express the result for the mass completely in terms of  $\rho_M$, $\rho_R$, and $R_0$ using
\begin{equation}
R_K'(r_0) = \pm \sqrt{1 - K R_0^2} =  \pm \sqrt{1 - R_0^2(\rho_M + \rho_R - 1)}
\end{equation}
where the minus sign corresponds to the case where the curvature is positive ($K = \rho_R + \rho_M - 1 > 0$) and the bubble includes more than half of the full sphere, $r_0 \sqrt{K} > \pi/2$.
Thus
\begin{equation}
\mu = R_H(R_H^2 + 1) = R_0^3 \rho_M \pm  2 \sqrt{\rho_R} R_0^2 \sqrt{1 - R_0^2(\rho_M + \rho_R - 1)} \; .
\end{equation}
In the positive curvature case where we have more than half of the sphere, there is a restriction $R_0 > 2 \sqrt{\rho_R}/\sqrt{(\rho_M + 2 \rho_R)^2 - 4 \rho_R}$ in order that the black hole has positive mass.\footnote{It is interesting that as this critical value of of $R_0$ is appoached from above, the solution degenerates into two disconnected parts, the exterior spacetime becoming pure AdS and the interior spacetime becoming a closed universe with a bubble of FRW spacetime glued together with a bubble of pure AdS spacetime.}
We can check that $R_0 > R_H$ in all cases;\footnote{The condition translates to the inequality $R_0(R_0^2 +1) \geq R_H(R_H^2 +1)$ or equivalently $R_0^2(1 - \rho_M) + 1 \ge \pm  2 \sqrt{\rho_R} R_0 \sqrt{1 - R_0^2(\rho_M + \rho_R - 1)}$. The left side of the latter is always positive, since for $\rho_M > 1$ we have $K > 0$ and then $R_0 < 1/\sqrt{K} = 1/\sqrt{\rho_M + \rho_R - 1} < 1/\sqrt{\rho_M - 1}$. The inequality then holds iff $(LS)^2 > (RS)^2$, which follows since $LS^2 - RS^2 = (R_0^2(1-\rho_M-2\rho_R) + 1)^2 > 0$.} this is required since $R_H$ is the smallest sphere radius on the $T=0$ slice of the Schwarzchild geometry.

\paragraph{Condition for bubbles behind the horizon}

The bubble cosmology will be completely behind the black hole horizon if and only if $\epsilon < 0$. Considering Equation (\ref{Junction2}) at $t=0$, we see that this is equivalent to
\begin{equation}
{R_K'(r_0) \over R_K(r_0)} < \sqrt{\rho_R} \; .
\end{equation}
This will be true if and only if
\[
\rho_M + 2 \rho_R > 1 \qquad {\rm and} \qquad r_0 > r_c 
\]
where $r_c$ is the smallest value of $r_0$ such that
\[
R_K(r_0) = 1/\sqrt{2 \rho_R + \rho_M - 1} \; .
\]

\paragraph{Condition for a complete causal patch}
If we want the cosmology to contain the entire causal patch of a comoving observer, we require that $r_0 > r_{obs}$, where $r_{obs}$ is the proper radius at $t=0$ of the forward light cone of a point on the big bang singularity. This light cone is described by a trajectory with $dt = a(t) dr$. Expressing $dt$ in terms of $da$ using the Friedmann equation, we have
\begin{equation}
\label{eq:robs}
r_{obs} = \int_0^1 {da \over \sqrt{\rho_M a + \rho_R - a^4 - (\rho_M + \rho_R - 1)a^2}} \; .
\end{equation}

\subsubsection*{Full Lorentzian solutions}

\begin{figure}
\centering
\includegraphics[width=0.29\linewidth]{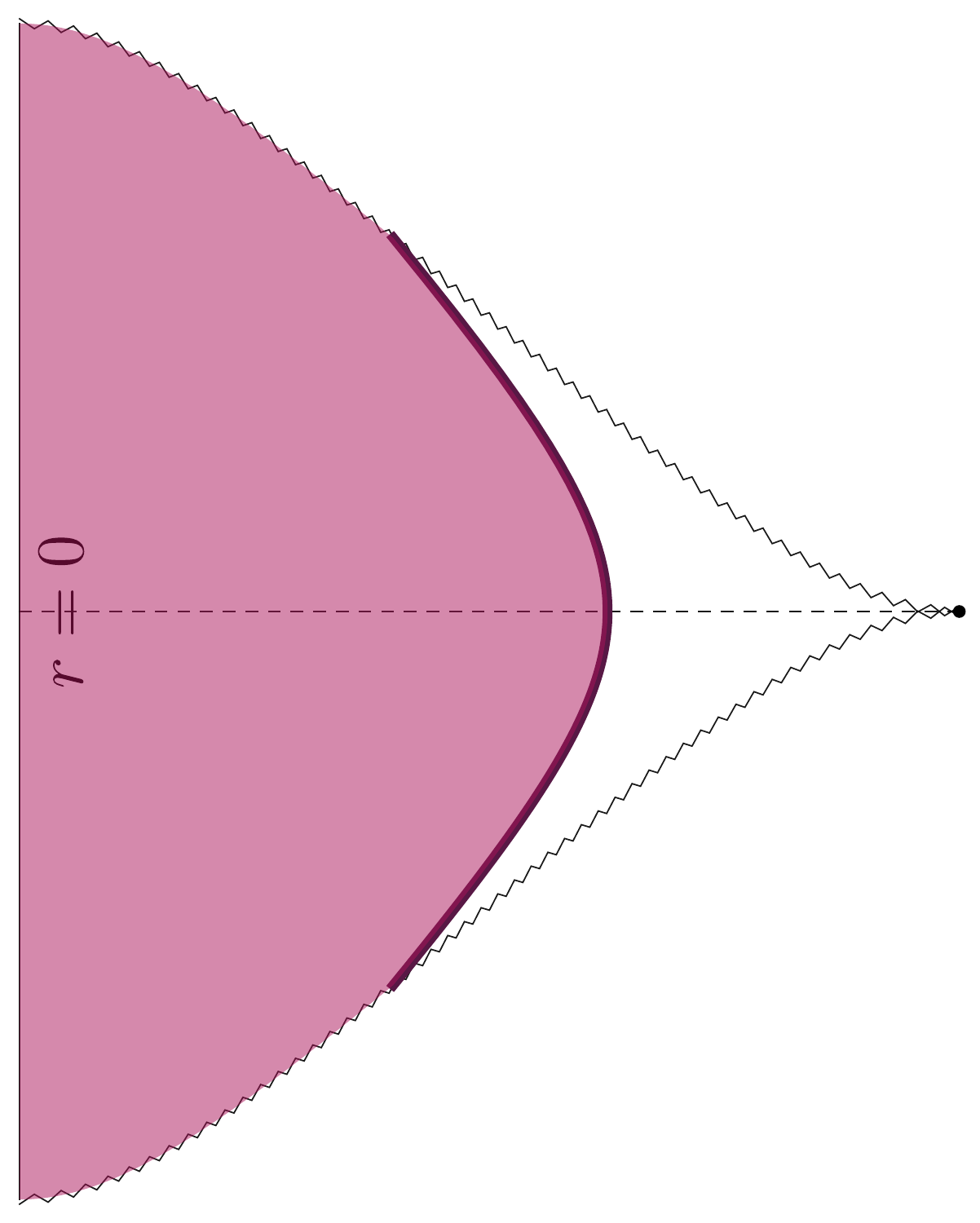}
\hspace{0.12\textwidth}
\includegraphics[width=0.34\linewidth]{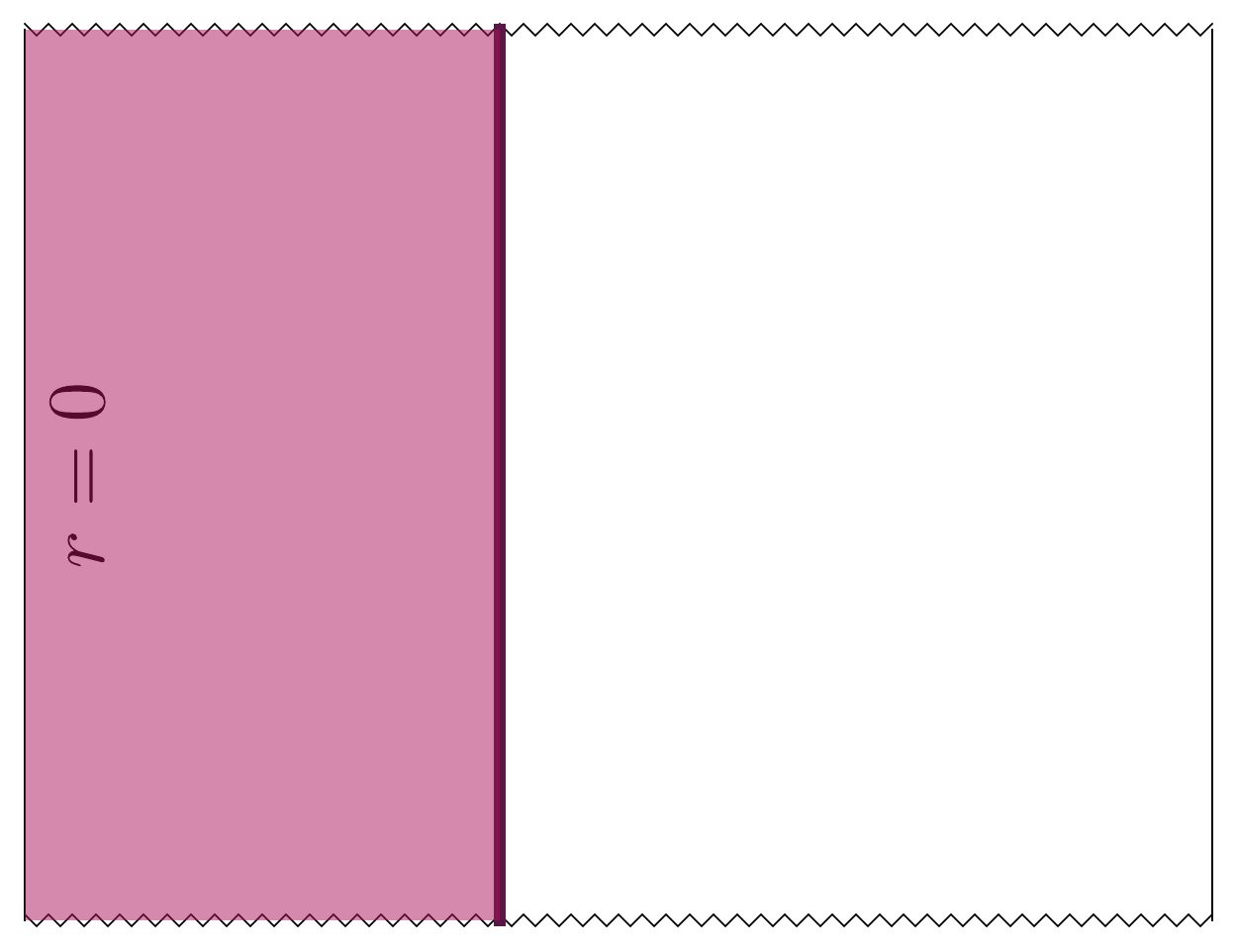}
\includegraphics[width=0.34\linewidth]{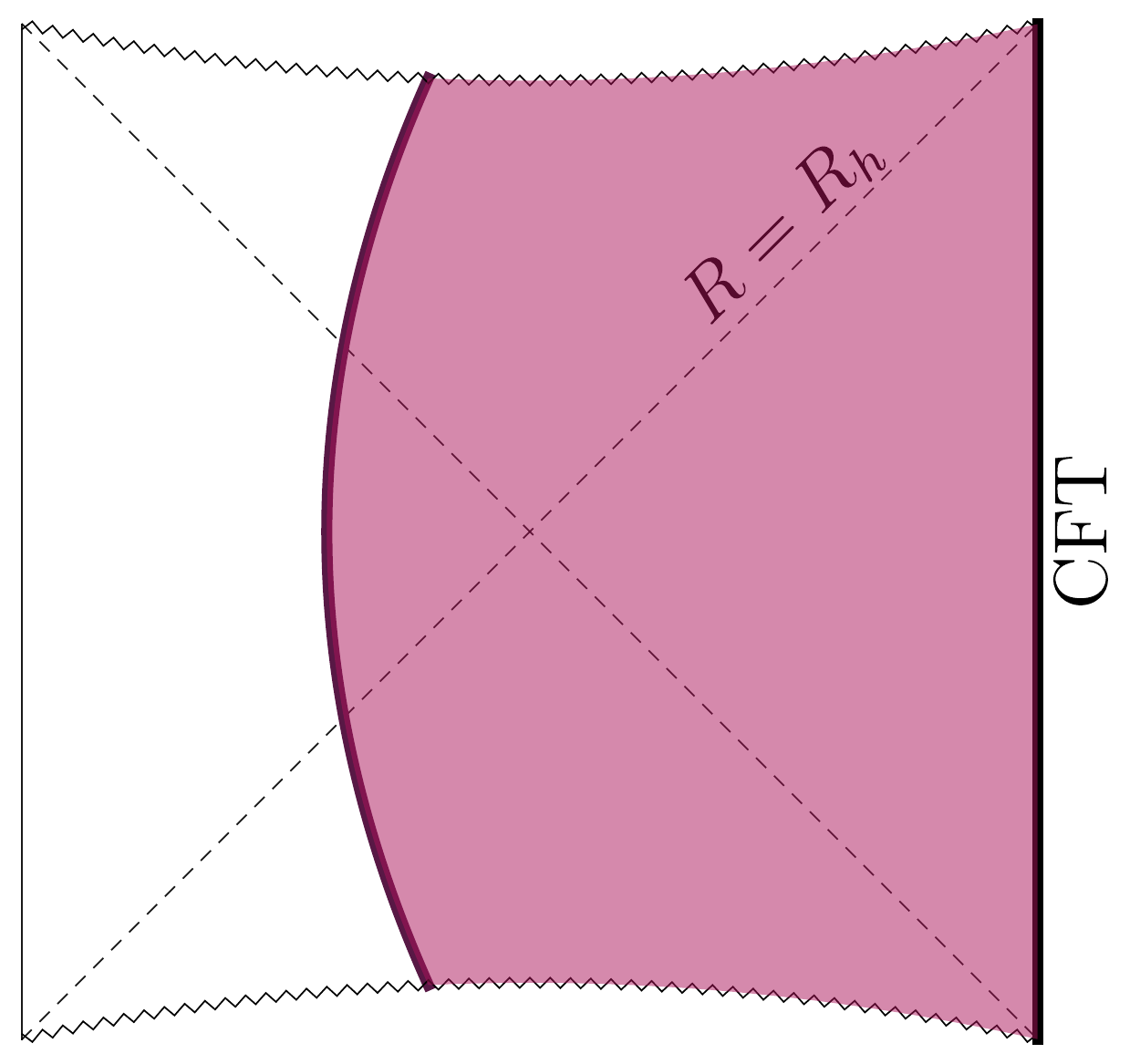} 
\hspace{0.12\textwidth}
\includegraphics[width=0.34\linewidth]{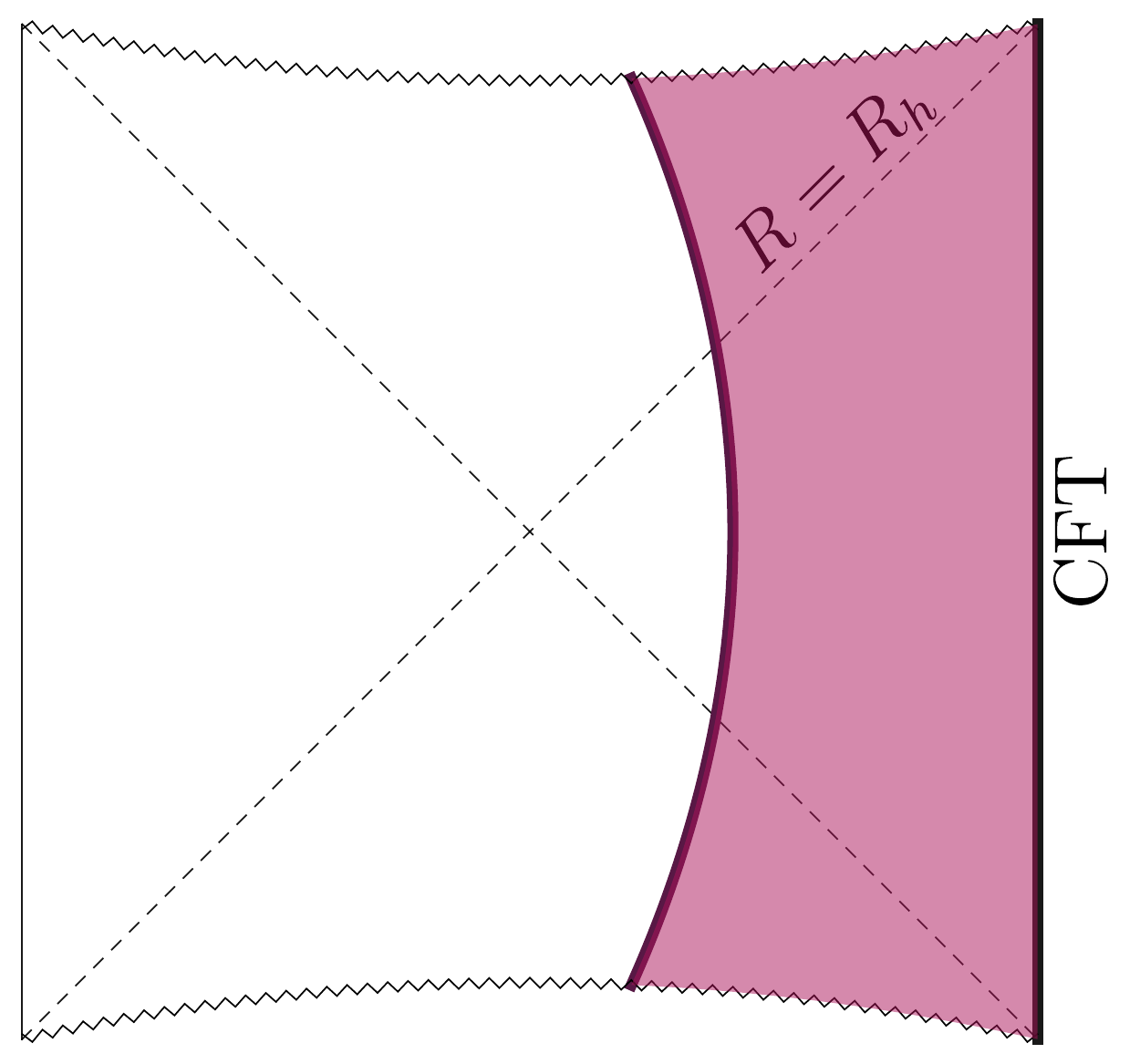} 
\caption{Lorentzian spacetime diagrams for solutions with an FRW bubble inside a Schwarzschild geometry. Top: shaded regions indicate spherical bubble of flat/open (left) and closed (right) $\Lambda < 0$ FRW cosmology, bounded by a comoving domain wall. Bottom: these bubbles are patched on part of an AdS-Schwarzschild spacetime indicated by the shaded region. The bubble may be completely inside the horizon (left) or begin outside the horizon (right).}
\label{fig:DWTrajectory}
\end{figure}

Since the bubble wall follows a comoving trajectory in the FRW region, the bubble geometry exists for finite proper time, starting from zero size at the big bang singularity in the past and ending at zero size at the big crunch singularity. 
In the Schwarzchild geometry, the bubble wall trajectory must also be timelike, with finite proper time, and from this perspective the interpretation is that the bubble starts at the past singularity of the black hole and ends at the future singularity of the black hole. 
If the bubble starts outside the horizon, it must cross the horizon at some finite proper time.
If the bubble starts inside the horizon, it remains inside the black hole at all times. The causal structures of the full Lorentzian solutions are depicted in Figure (\ref{fig:DWTrajectory}). 

\subsection{Examples}

It will be useful to have some explicit examples. 

\paragraph{Pure radiation}

For pure radiation, we can solve the Friedmann equation explicitly to find
\begin{equation}
    a(t) = \sqrt{(1 + \rho_R) \cos^2(t) - \rho_R} \; \qquad \qquad {\rm  pure \; radiation}.
\end{equation}
The spatial curvature is $K=\rho_R - 1$, so $\rho_R = 1$ gives the special case of flat cosmology, where $a(t) = \sqrt{\cos(2t)}$.

From (\ref{eq:robs}), we find that the size of a bubble corresponding to the causal diamond of a single observer is\footnote{Here $K(x)$ is the elliptic integral that approaches $\pi/2$ for small $x$ and $\ln(8/(1-x))/2$ as $x \to 1$.}
\begin{equation}
    r_{obs} = {1 \over \sqrt{\rho_R + 1}}K \left({1 \over \sqrt{\rho_R + 1}} \right) \qquad \qquad {\rm  pure \; radiation}
\end{equation}
In the flat case $\rho_R = 1$, we have 
\[
r_{obs} = \sqrt{\pi} \Gamma(5/4)/\Gamma(1/4) \approx 1.311 \qquad \qquad {\rm flat, \; pure \; radiation} \; .
\]

\paragraph{Pure matter}

For bubbles of pure matter cosmology, the black hole mass parameter $\mu$ is simply
\begin{equation}
    \mu = R_0^3 \rho_M \; 
\end{equation}
In the flat case, this equal to the total mass of matter inside the bubble.

For a flat cosmology with pure matter, we have $\rho_M = 1$ and the Friedmann equation can be solved explicitly to give
\begin{equation}
   a(t) = \cos^{2 \over 3}\left( {3 \over 2} t \right) \qquad \qquad {\rm flat, \; pure \; matter}
\end{equation}
Here, the proper radius of the causal patch accessible to a single observer is 
\[
r_{obs} = 2 \sqrt{\pi} \Gamma(7/6)/\Gamma(2/3) \approx 2.43 \qquad \qquad {\rm flat, \; pure \; matter}\; .
\]

\section{Thermodynamics}

We have seen that it is possible to construct a solution for which a bubble of $\Lambda < 0$ cosmology with arbitrary matter and radiation densities and arbitrary size is embedded in an asymptotically AdS Schwarzchild geometry. We would like to understand whether such solutions can correspond to legitimate states of a CFT.

An initial worry is a sharp version of the well-known bag-of-gold puzzle. For cosmologies with radiation, we can calculate the entropy of this radiation and compare it with the Bekenstein-Hawking entropy of the black hole that contains the cosmological bubble. If the black hole entropy is smaller, then the putative dual CFT cannot accommodate all of the states of the radiation as independent states. It's still possible that these bulk states are encoded non-isometrically; this would imply that certain cosmological bubble states can be expressed as linear combinations of other such states.

The Bekenstein-Hawking entropy of the black hole is
\begin{equation}
    \label{eq:SBH}
    S_{BH} = {A_H \over 4 G_N} = {\pi R_H^2 \over G_N} \; .
\end{equation}
Below, it will also be useful to have the black hole temperature and energy
\begin{equation}
    T = {1 + 3 R_H^2 \over 4 \pi R_H} \qquad \qquad E = {1 \over 2 G_N} (R_H + R_H^3) \; .
\end{equation}
The volume of the bubble of cosmology can be calculated as
\begin{equation}
\label{eq:volume}
   V_K(r_0) = \int_0^{r_0} 4 \pi R^2_K(r) dr = \left\{
\begin{array}{ll} 
\pi  K^{-{3 \over 2}} \left( 2 r_0\sqrt{K} - \sin(2 r_0 \sqrt{K} )\right) & K > 0 \cr
 {4 \over 3} \pi r_0^3 & K=0 \cr
\pi |K|^{-{3 \over 2}} \left(\sinh(2  r_0\sqrt{|K|}) -2  r_0\sqrt{|K|}   \right) & K < 0
\end{array}
\right.
\end{equation}
Assuming an order one number of light fields in the cosmology and ignoring factors of order one, the entropy density $s$ of radiation is related to the energy density $u$ by $s \sim u^{3/4}$. We defined $\rho_R$ as $8 \pi G_N /3$ times the energy density, so the entropy of radiation is
\[
S_{rad} \sim V \left( {\rho_R \over G_N} \right)^{3 \over 4} \; .
\]
If $r_0$, $\rho_M$, and $\rho_R$ are all of the scale set by the cosmological constant, the black hole entropy will be much larger, $S_{BH}/S_{rad} \sim \sqrt{\ell_{AdS}/\ell_P} \gg 1$. In particular, this will hold for bubbles containing the full region accessible to a single observer, but also much larger bubbles that are not parametrically larger.\footnote{We do have a special case for $K > 0$ when considering bubbles containing more than half of the sphere. Here, the black hole mass and entropy decrease as $r_0$ increases and go to zero at $R_0 = 2 \sqrt{\rho_R}/\sqrt{(\rho_M + 2 \rho_R)^2 - 4 \rho_R}$. At this point, the spacetime pinches off into pure AdS outside and a disconnected closed universe with a cosmological bubble joined to a bubble of pure AdS. Here, the cosmology is clearly not described by a dual CFT state.}

If we consider bubbles much larger than the cosmological scale (for the $K \le 0$ cases), we have for $K < 0$ that 
\[
R_H \approx R_0\left(\rho_M + 2 \sqrt{\rho_R(1 - \rho_M - \rho_R)}\right)^{1 \over 3}
\]
In this negatively curved case, the volume behaves for large $R_0$ like $R_0^2/ \sqrt{K}$, so the black hole entropy and the radiation entropy both scale as $R_0^2$ for large $R_0$. In this case, the black hole entropy remains larger than the radiation entropy by a factor of $\sqrt{\ell_{AdS}/\ell_P}$ for bubbles of all size. 

In the flat case, we can have sufficiently large bubbles such that an entropy puzzle arises. Here, for large $R_0$, we have $V \sim R_0^3$. If there is matter, we find for large $R_0$ that
\[
R_H \approx R_0 \rho_M^{1 \over 3}
\]
and the entropy of the radiation becomes larger than the black hole entropy when 
\[
{R_0 \over \ell_{AdS}} \gtrsim {\rho_M^{2 \over 3} \over \rho_R^{3 \over 4}} \sqrt{ \ell_{AdS} \over \ell_P} \; .
\]
For a flat universe with radiation only, we have for large $R_0$ that 
\[
R_H \approx R_0^{2 \over 3} \rho_R^{1 \over 6} 
\]
In this case the entropy of the radiation will exceed the black hole entropy when
\begin{equation}
    {R_0 \over \ell_{AdS}} \gtrsim \left({ \ell_{AdS} \over \ell_P}\right)^{3 \over 10}\; .
\end{equation}
Both of these critical bubble sizes would become infinite in the usual large $N$ limit of the CFT where $\ell_{AdS} / \ell_P \to \infty$.

In both of these cases, the geometries would correspond to extremely large CFT energies. In units of the CFT sphere radius, the energies are $E \gtrsim \rho_M^3/\rho_R^{9/4}(\ell_{AdS} / \ell_P)^{7/2}$ in the case with matter and radiation
and $E \gtrsim (\ell_{AdS} / \ell_P)^{13/5}$ in the pure radiation case. 

In general, we should also expect some interior entropy associated with the matter, but the details of this will depend on the construction. In situations where a significant fraction of the energy density is in radiation, we would expect the matter entropy to be smaller. For pure matter cases, it could be interesting to perform an analysis similar to that of \cite{Balasubramanian:2022gmo}, replacing the shells of matter with the approximately uniform ball of matter that we are considering here.

\section{Euclidean continuations}

We have constructed asymptotically AdS geometries containing a bubble of $\Lambda < 0$ cosmology with arbitrary initial matter densities $\rho_M$ and $\rho_R$ and arbitrary radius $r_0$ (subject to the constraint of footnote 4 in the positive curvature case). 
In order to understand better whether these can correspond to some legitimate states of a dual CFT, we will consider in this section the possibility of constructing such states directly via a Euclidean path integral construction. Path integral states describing bubbles of pure AdS cosmology inside asymptotically AdS spacetimes with a different cosmological constant were constructed in \cite{fu_bag--gold_2019,simidzija_holo-ween_2020}. On the other hand, \cite{fu_bag--gold_2019} found that geometries with a bubble of de Sitter cosmology inside AdS appear not to have a path-integral construction.

Pure global AdS spacetime corresponds to the CFT vacuum state, which can be constructed via the Euclidean path integral on an infinite cylinder $S^2 \times \mathbb{R}$ where the $\mathbb{R}$ direction represents Euclidean time. This CFT path integral corresponds to a gravity path integral whose saddle-point geometry is pure global AdS. By inserting operators or sources into this CFT path integral, we can construct other states. To produce time-reversal invariant Lorentzian geometries, we can choose sources in the Euclidean picture to be symmetric under Euclidean time reversal. The Lorentzian geometry can be obtained by determining the dual Euclidean geometry via the rules of AdS/CFT (where the sources and operator insertions determine the boundary conditions for the Euclidean spacetime) and then performing an analytic continuation $\tau \to it$ of the Euclidean time coordinate, where $t=\tau = 0$ represents the time-reversal / reflection invariant slice.

Since we are starting from a set of Lorentzian geometries, we will try to apply this procedure in reverse, first performing an analytic continuation to Euclidean geometries and then (if these are sensible) looking at the asymptotic behaviour to understand what types of insertions/sources are required in the dual CFT path integral.

\paragraph{FRW part}

The FRW part of the spacetime continues to a geometry
\begin{equation}
    ds^2 = d \tau^2 + a_E^2(\tau) (dr^2 + R_K(r) d \Omega^2)
\end{equation}
with boundary $r(\tau) = r_0$. Here, the Euclidean scale factor is defined via $a_E(\tau) = a(i \tau)$ and satisfies the analytically continued Friedmann equation
\begin{equation}
\left({a' \over a}\right)^2 = -{\rho_M \over a^3} - {\rho_R \over a^4} + {K \over a^2} + 1 \equiv H(a)^2 \; .
\end{equation}
In the Euclidean picture, the scale factor grows away from $\tau = 0$ so that the cosmological constant term comes to dominate and we have $a_E(\tau) \sim e^{\pm \tau}$ for $\tau \to \pm \infty$. We thus have asymptotically AdS regions for  Euclidean times $\tau \to \pm \infty$; for this reason, this solution is sometimes described as an asymptotically AdS Euclidean wormhole. 

\begin{figure}
    \centering
    \includegraphics[width=0.32\linewidth]{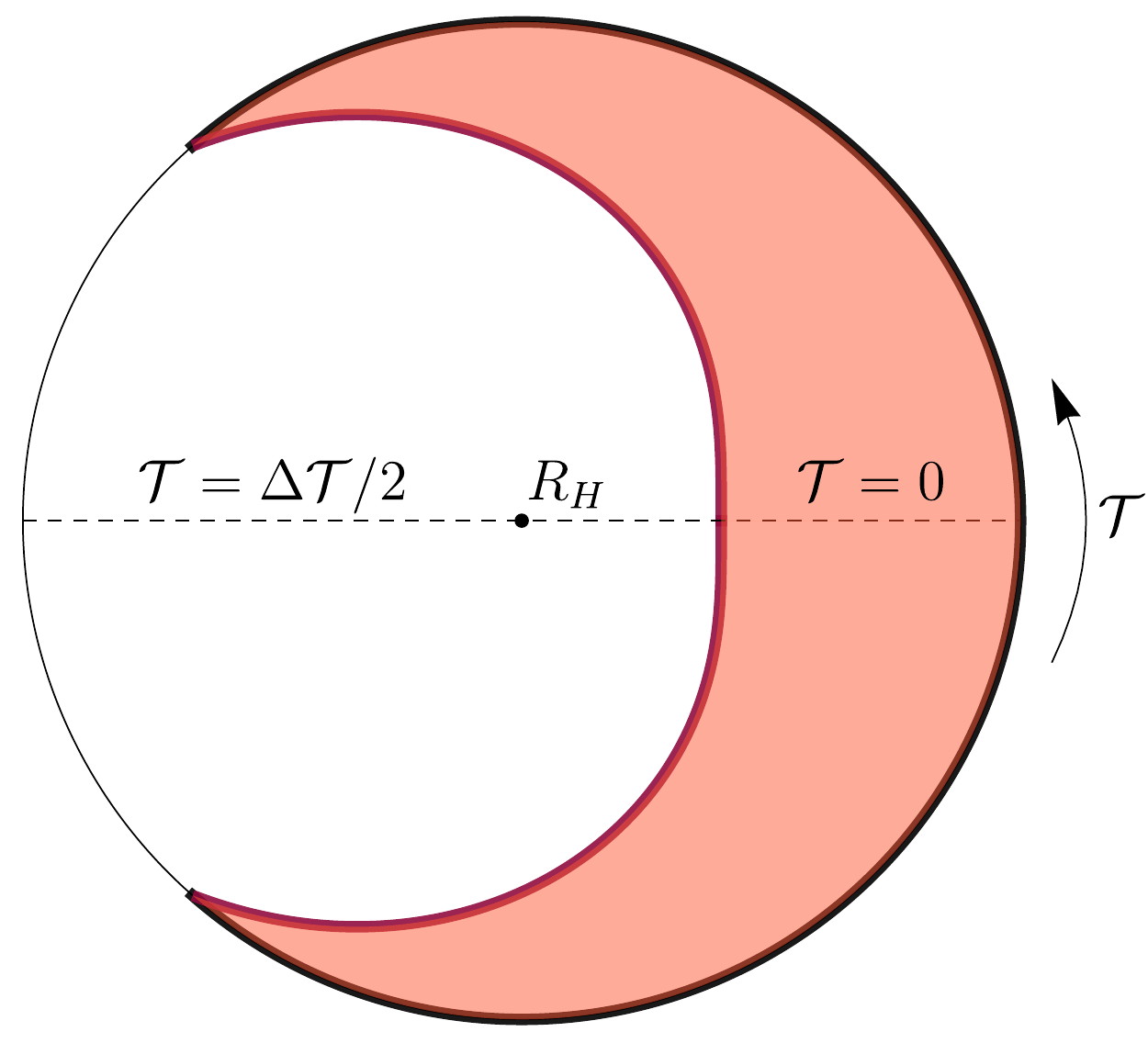}
     \includegraphics[width=0.32\linewidth]{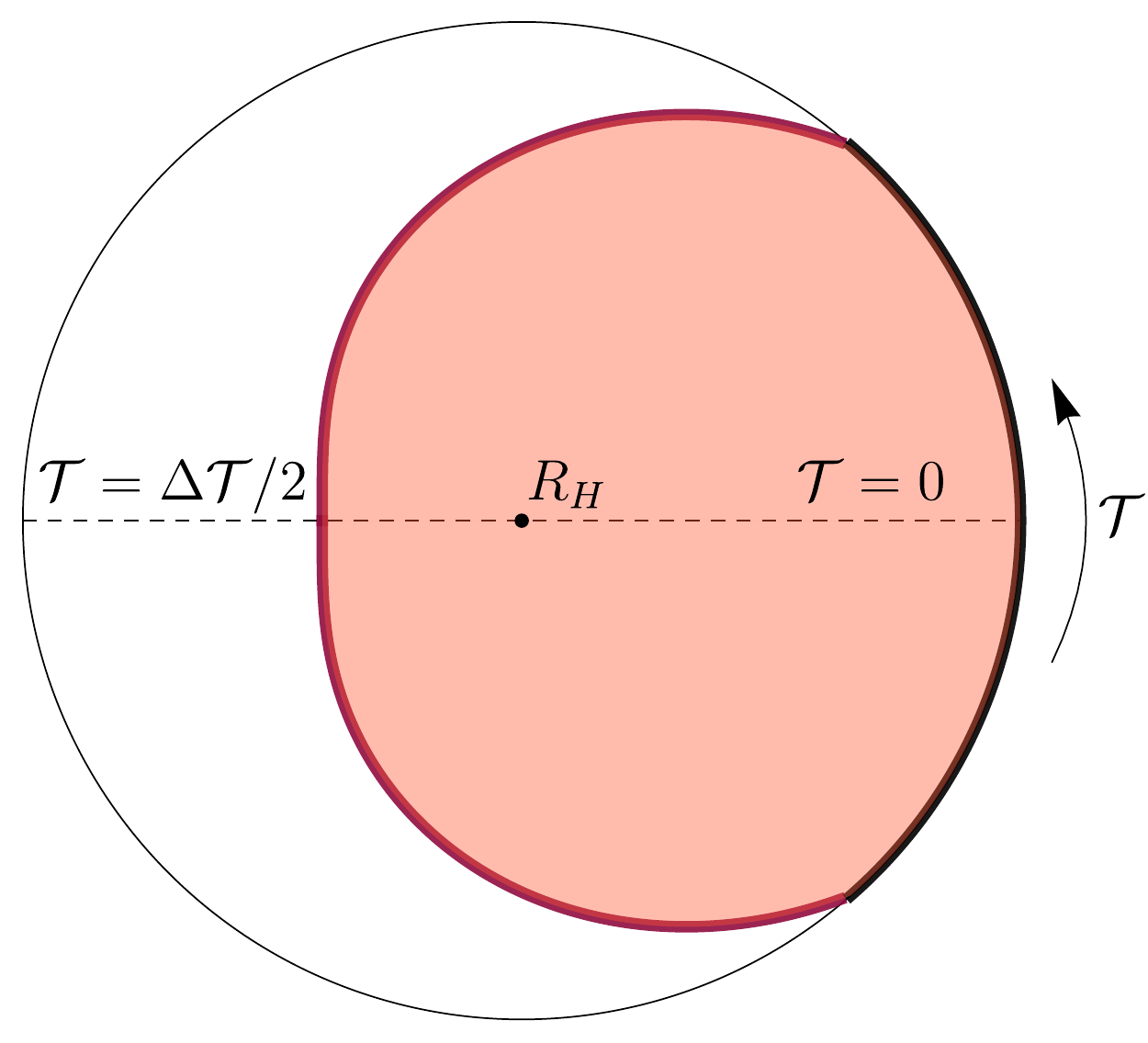}
    \includegraphics[width=0.32\linewidth]{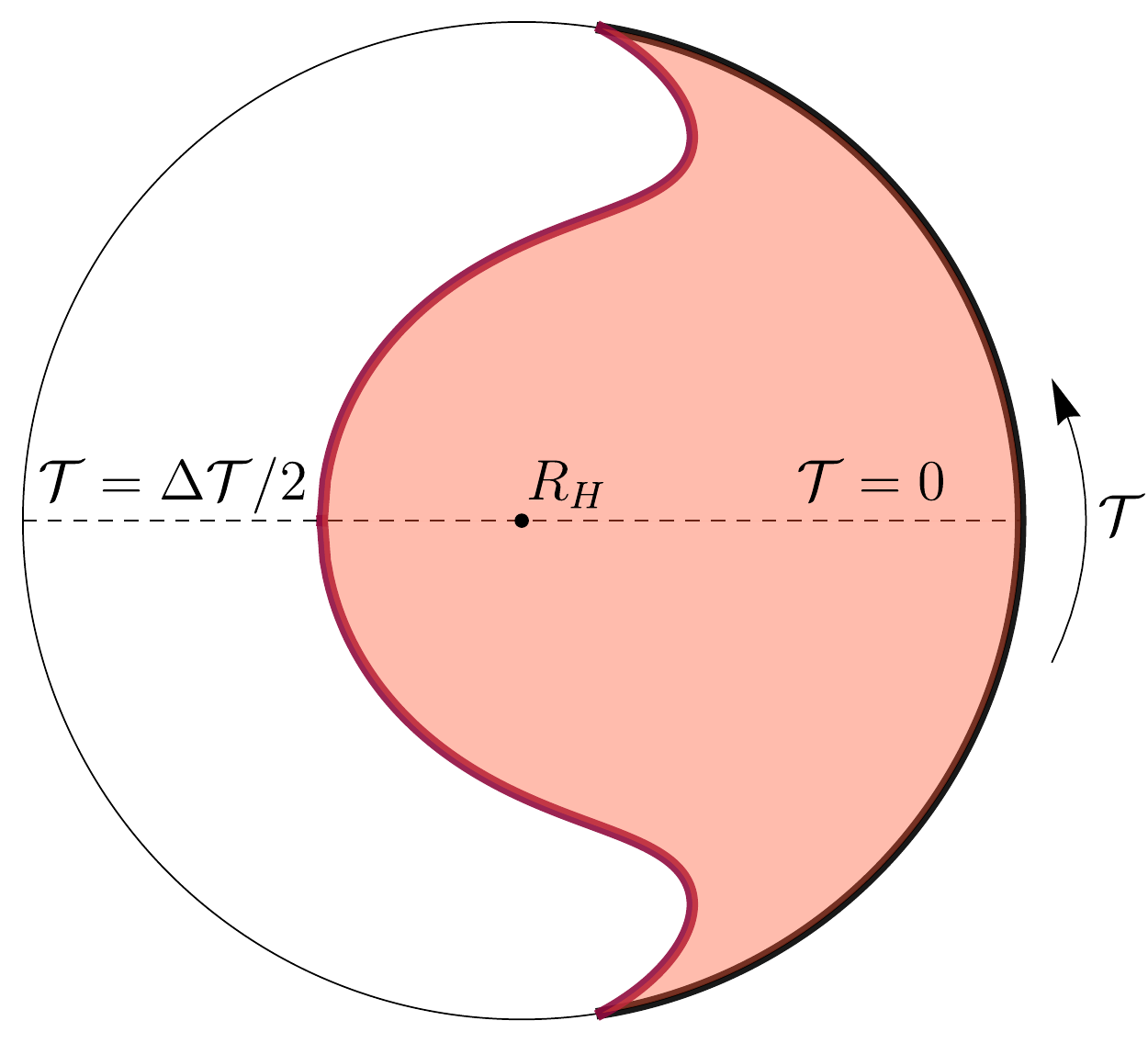}
    \includegraphics[width=0.32\linewidth]{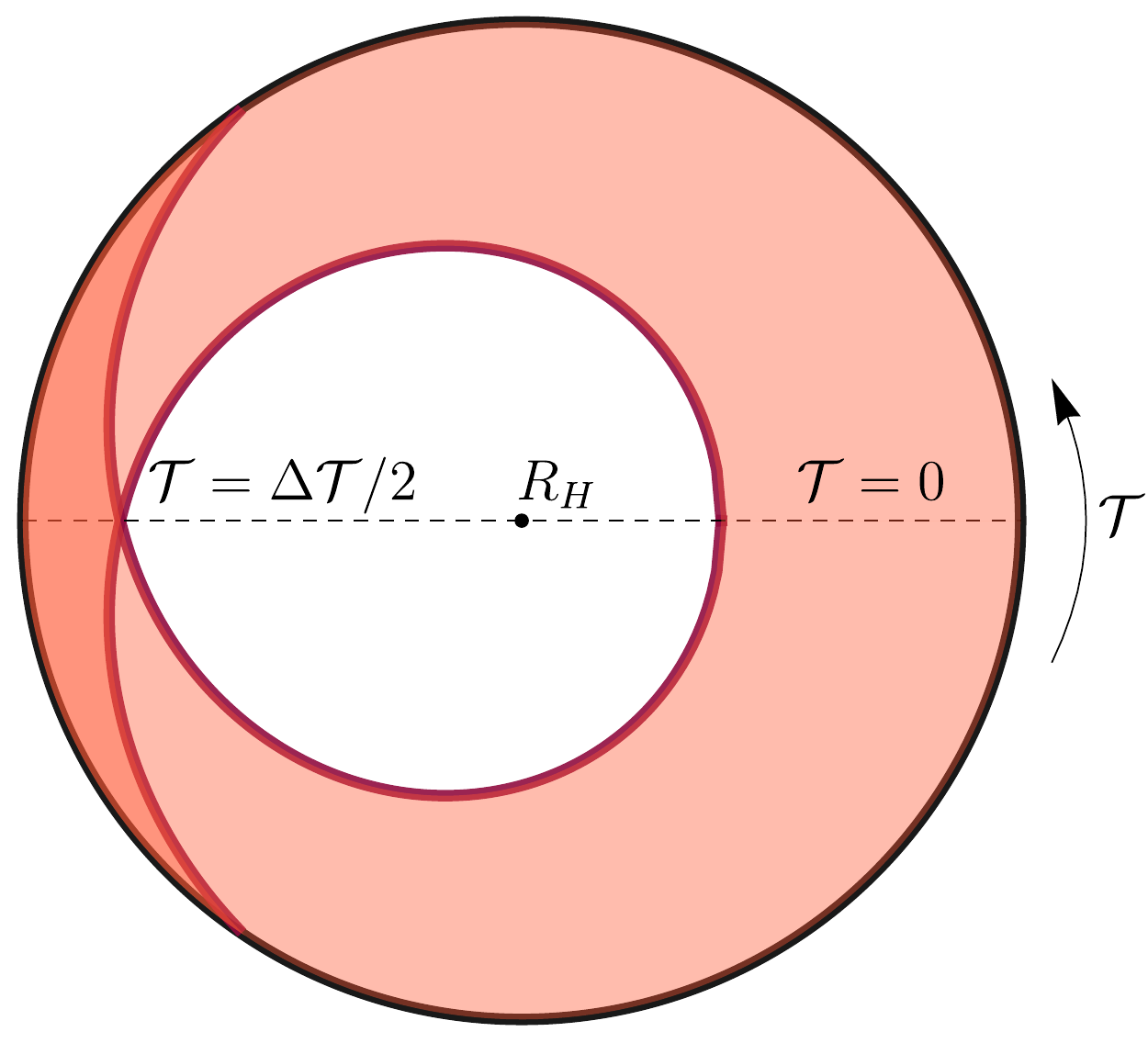}
    \includegraphics[width=0.32\linewidth]{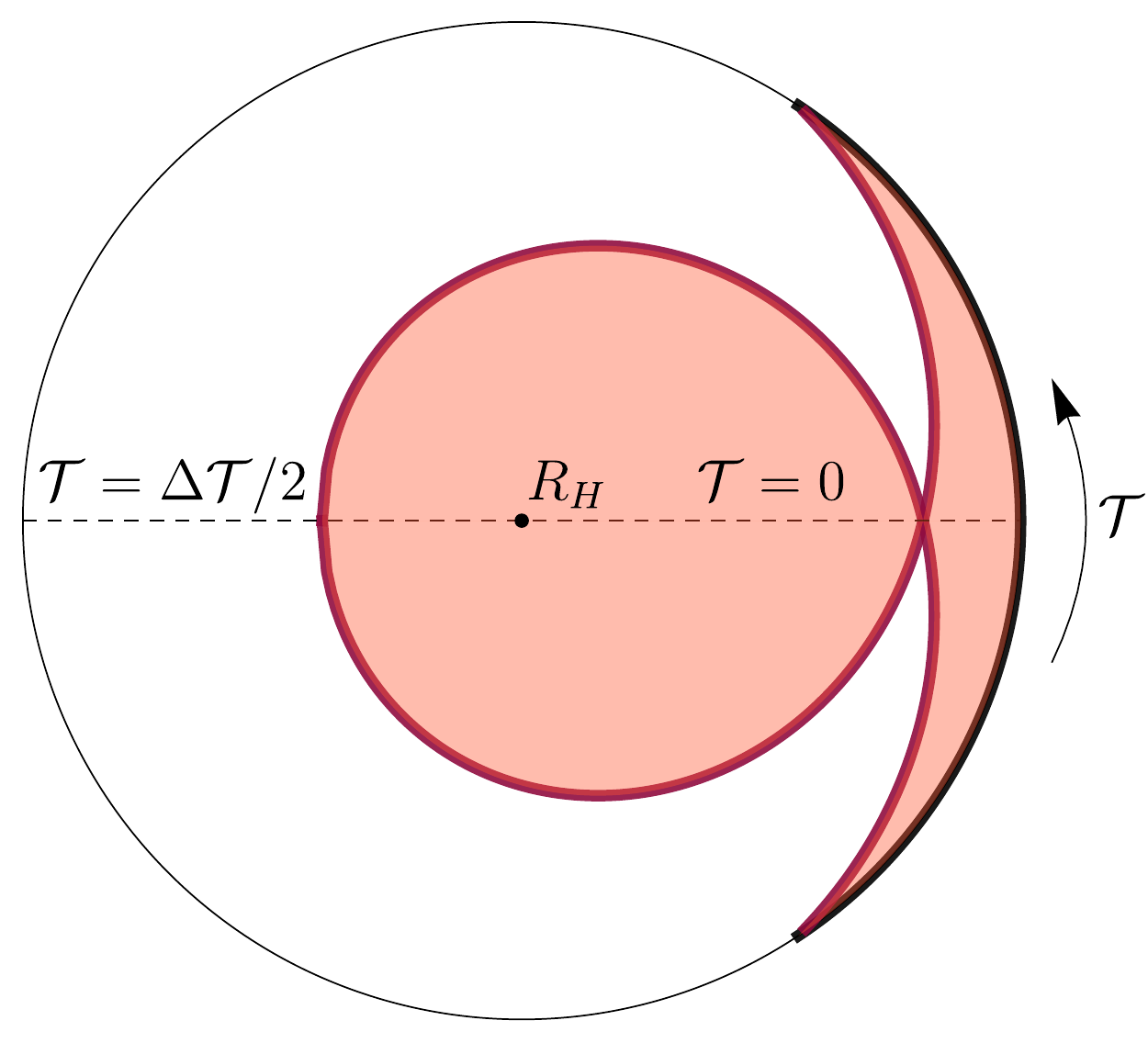}
    \includegraphics[width=0.32\linewidth]{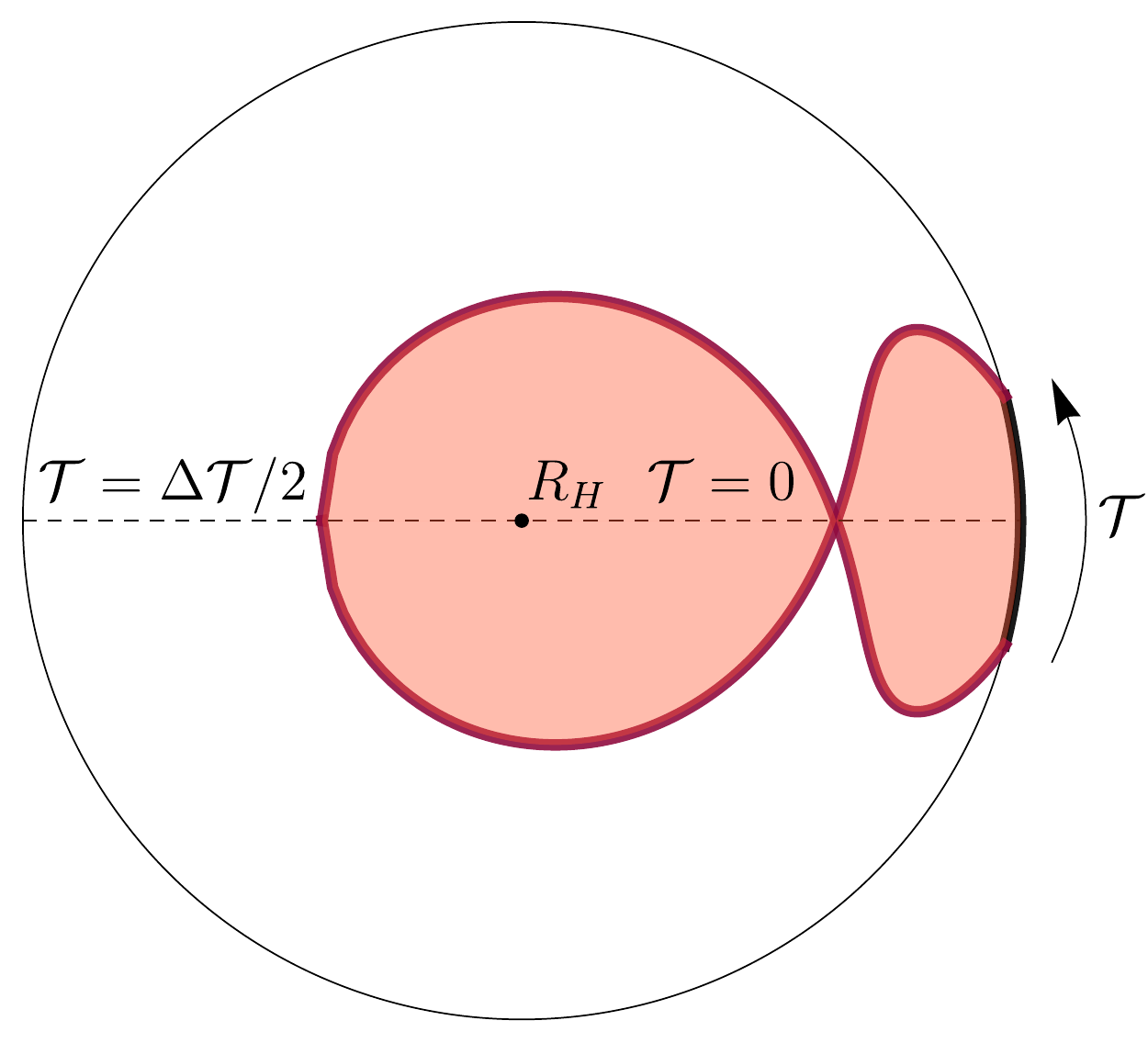}
    \caption{Possible domain wall trajectories (thick maroon lines) in the Euclidean Schwarzschild spacetime with the shaded region corresponding to the part that is glued to the Euclidean FRW bubble. The first column corresponds to spacetimes where the domain wall is initially outside the horizon in the Lorentzian solution. In the bottom row, the bubble wall self-intersects in the second and third examples and the spacetime is not sensible. The first example on the bottom is sensible as it corresponds to a part of the covering space with ${\cal T} \in (-\infty,\infty)$.
  }
    \label{fig:EucDWCases}
\end{figure}

\paragraph{Schwarzschild exterior}The exterior region forms a part of the Euclidean AdS-Schwarzschild geometry. This can be described as 
\begin{equation}
ds^2 = {dR^2 \over F(R)} + F(R) d {\cal T}^2 + R^2 d \Omega^2 \qquad \qquad F(R) = R^2 + 1 - {\mu \over R}\; ,
\end{equation}
with $F(R)$ as before. We have $F(R) = 0$ at $R=R_H$. If this point is part of our solution (required when the bubble is fully inside the black hole), it is necessary that ${\cal T}$ is taken to be periodic with period
\begin{equation}
    \Delta {\cal T} = {4 \pi R_H \over 1 + 3 R_H^2}
\end{equation}
to avoid a conical singularity.
Matching the induced metric from the the FRW region with the induced metric from the Schwarzchild region, the bubble trajectory in this Euclidean spacetime must satisfy
\begin{equation}
    R(\tau) = a_E(\tau) R_0 \equiv a_E(\tau) R_K(r_0) \qquad {R'(\tau)^2 \over F(R(\tau))} + F(R(\tau))({\cal T}'(\tau))^2 = 1 \; ,
\end{equation}
the analytically continued version of (\ref{eq:SchBubble}). It is convenient to parametrize the trajectories in terms of $a_E$ (which we will write as $a$ going forward). In this case, we have
\begin{equation}
R(a) = R_0 a \qquad {d {\cal T} \over da} = \pm {1 \over F(R_0 a) a H(a)} \sqrt{ F(R_0 a)  - R_0^2 a^2 H^2(a)}
\end{equation}
The square root here is precisely the one appearing in the analytically continued version of (\ref{Junction2}), so we can rewrite the ${\cal T}$ equation as
\begin{equation}
\label{eq:dTda}
{d {\cal T} \over da} =  {1 \over F(R_0 a) a H(a)} \left(R_1 - \sqrt{\rho_R} {R_0 \over a}\right)
\end{equation}
where $R_1 \equiv R_K'(r_0) $. The expression in brackets is exactly the quantity that is positive if bubble is initially outside the horizon and negative otherwise. 

\paragraph{Bubble initially outside horizon}

If the bubble is initially outside the black hole horizon, the expression (\ref{eq:dTda}) is initially positive (when $a=1$) and therefore positive on the entire trajectory, since $a \ge 1$ in the Euclidean solution. We can take the region $\{{\cal T} = 0, R > R_0\} $ of the Schwarzschild geometry as the initial slice for the part of the solution we are keeping (leftmost figures in \ref{fig:EucDWCases}). The boundary of the region we keep is determined by solving (\ref{eq:dTda}). 

The ${\cal T}$ coordinate increases from 0 to some maximum value ${\cal T}_0$ with
\begin{equation}
    {\cal T}_0 = \int_1^\infty {da \over F(R_0 a) a H(a)} \left(R_1 - \sqrt{\rho_R} {R_0 \over a}\right)
\end{equation}
Since the solution in this case does not include the Euclidean horizon (the point $R = R_H$), there is no restriction on how large the range can be (i.e. we do not require $2{\cal T}_0 \le \Delta {\cal T}$).

\paragraph{Bubble initially inside horizon}

When the cosmological bubble is fully inside the horizon, we can take the region $\{{\cal T} = 0\} \cup \{{\cal T} = \Delta {\cal T}/2, R < R_0\} $ of the Schwarzschild geometry as the initial slice for the part of the solution we are keeping (four figures on the right in \ref{fig:EucDWCases}). 

In this case, the ${\cal T}$ coordinate initially decreases. If $R_1 < 0$ (positive curvature with more than half of the sphere), ${\cal T}'$ remains negative for all time. If $R_1 > 0$, ${\cal T}'$ will initially be negative, but drops to zero when $a = a_0 \equiv \sqrt{\rho_R} R_0 / R_1$ and switches sign afterwards.

\begin{figure}
    \centering    
    \includegraphics[width=0.45\linewidth]{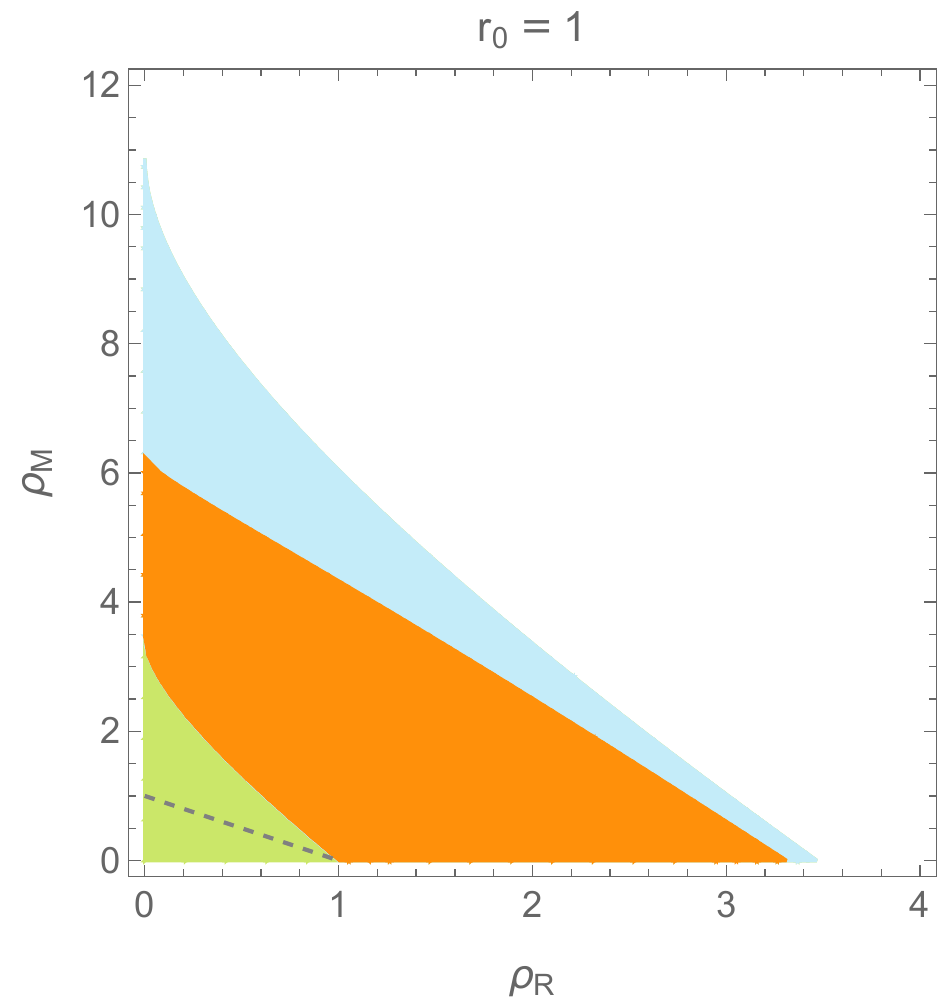}
    \includegraphics[width=0.45\linewidth]{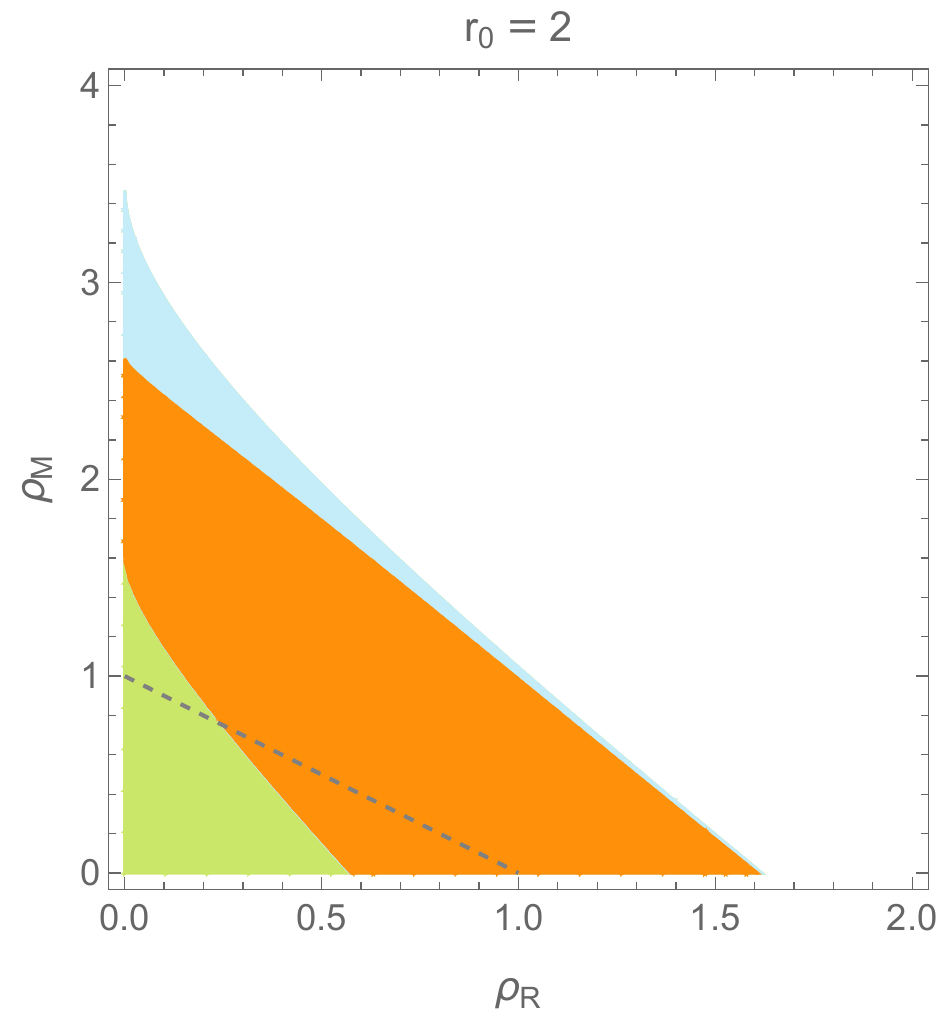}
    \includegraphics[width=0.45\linewidth]{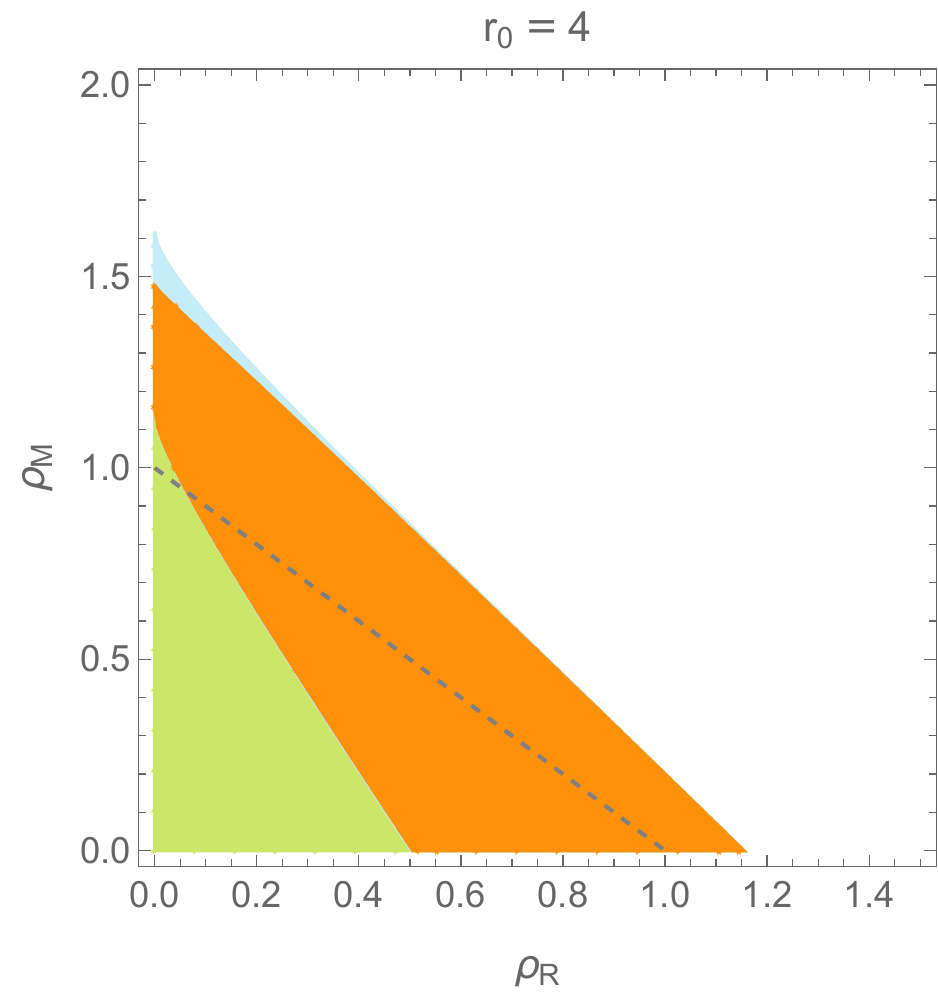}
    \includegraphics[width=0.45\linewidth]{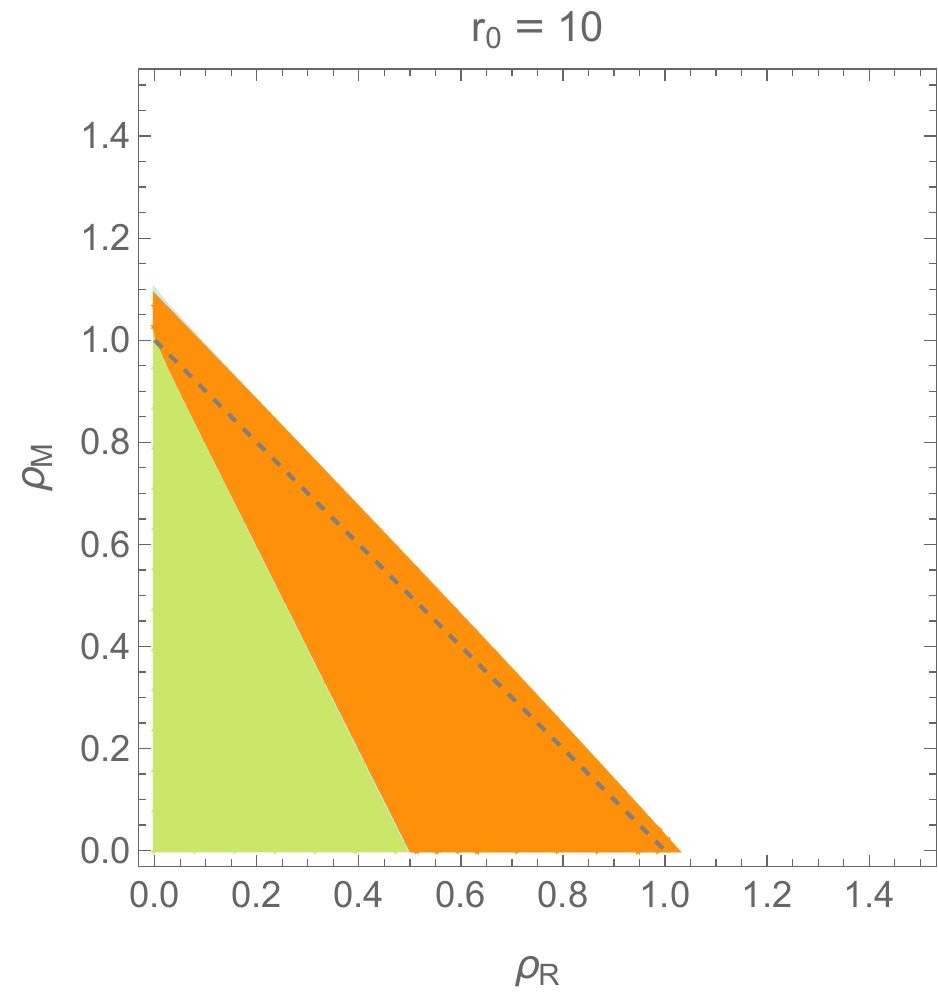}
    \caption{
    The shaded regions are parameter values for which the Lorentzian solution has a positive mass black hole. The Euclidean solution is sensible (the domain wall does not self-intersect) in the orange and green regions. The cosmological bubble is initially outside the horizon in the green region. The dashed line corresponds to $K=0$; below it $K<0$ and above it $K>0$.
    }
    \label{fig:good_asymptotics_region_plot}
\end{figure}

In either case, the range of ${\cal T}$ on the Euclidean boundary corresponding to the part of the geometry we are keeping is $[-{\cal T}_0,{\cal T}_0]$
\begin{equation}
    {\cal T}_0 = {\Delta {\cal T} \over 2} + \int_1^\infty {da \over F(R_0 a) a H(a)} \left(R_1 - \sqrt{\rho_R} {R_0 \over a}\right) \; .
\end{equation}
For the geometry to make sense, we require that the domain wall is not self-intersecting. In the $R_1 < 0$ case, this requires that ${\cal T}_0 > 0$. In the other case when ${\cal T}'$ switches sign at $a = a_0$, we require that 
\begin{equation}\label{eq:good_Euc_asymptotics}
    {\Delta {\cal T} \over 2} + \int_1^{a_0} {da \over F(R_0 a) a H(a)} \left(R_1 - \sqrt{\rho_R} {R_0 \over a}\right) > 0  \; .
\end{equation}
The region of parameter space for which the geometries with the bubble initially inside the horizon have good Euclidean asymptotics (i.e. the domain wall does not self intersect) are shown in Figure \ref{fig:good_asymptotics_region_plot}

Exploring these conditions numerically, we find that for $K \le 0$ where the spatial geometry is infinite, the solutions make sense for arbitrarily large bubbles. For cases with $K > 0$ (positively curved spatial geometry), we find that there is generally some $r_{int}$ above which the boundary self-intersects in the Euclidean solution. As far as we can tell, this is always larger than half the sphere diameter $r_{max} = \pi/\sqrt{K}$ but smaller than the value $r_{pinch}$ at which the exterior black hole mass goes to zero. 

\begin{figure}
    \centering    
    \includegraphics[width=0.45\linewidth]{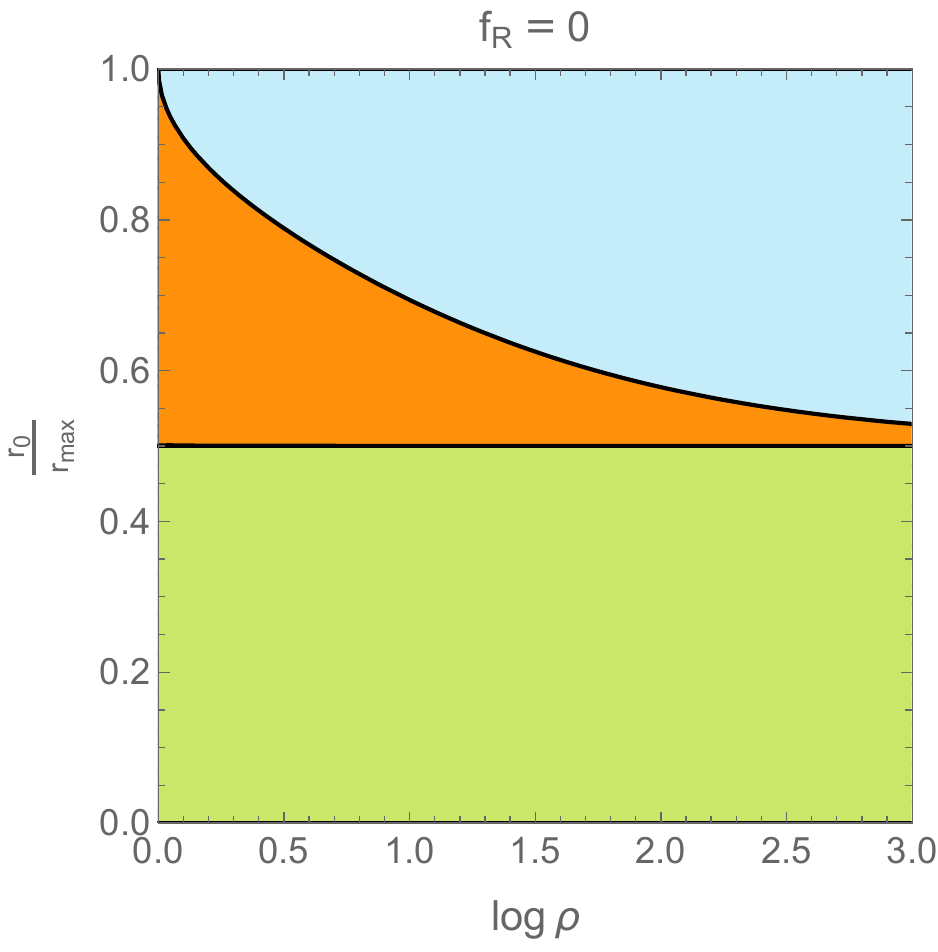}
    \includegraphics[width=0.45\linewidth]{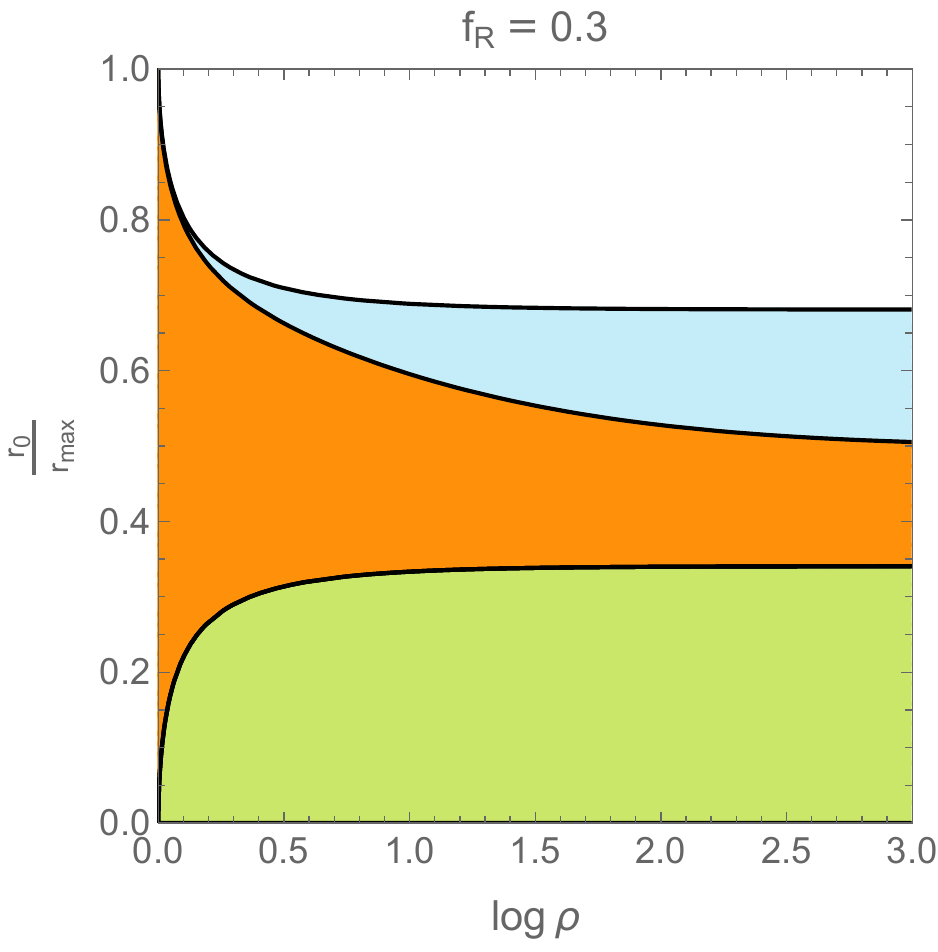}
    \includegraphics[width=0.45\linewidth]{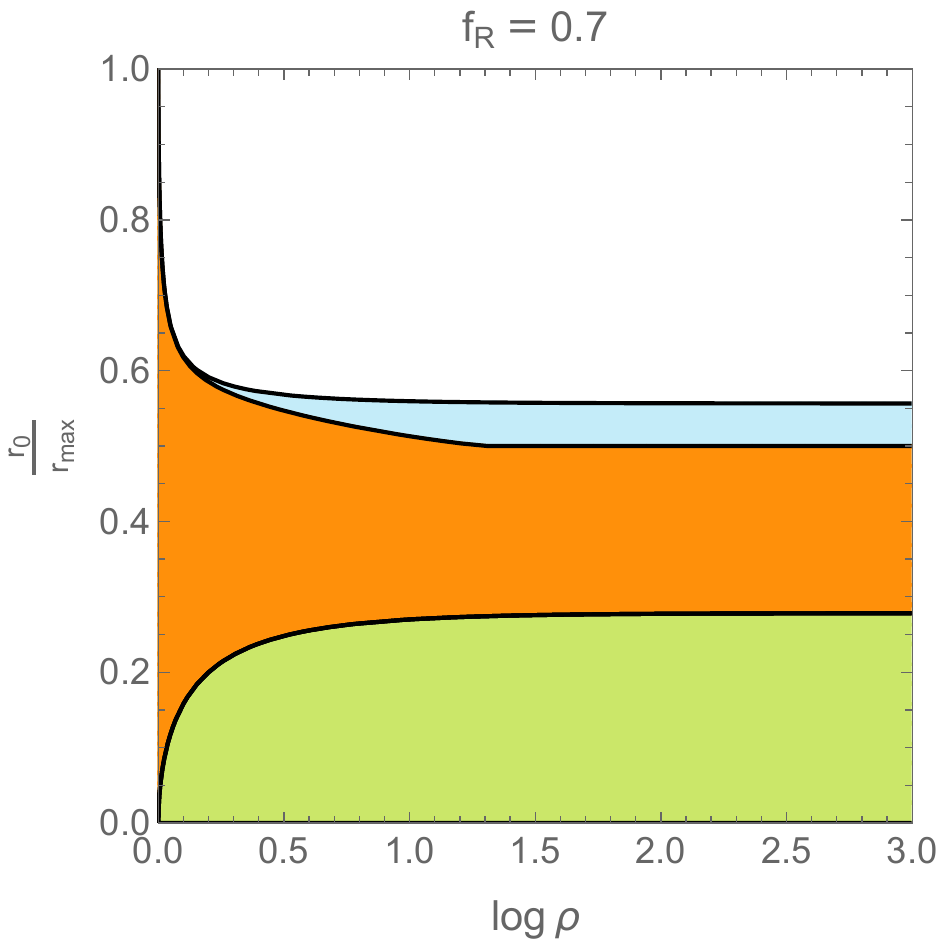}
    \includegraphics[width=0.45\linewidth]{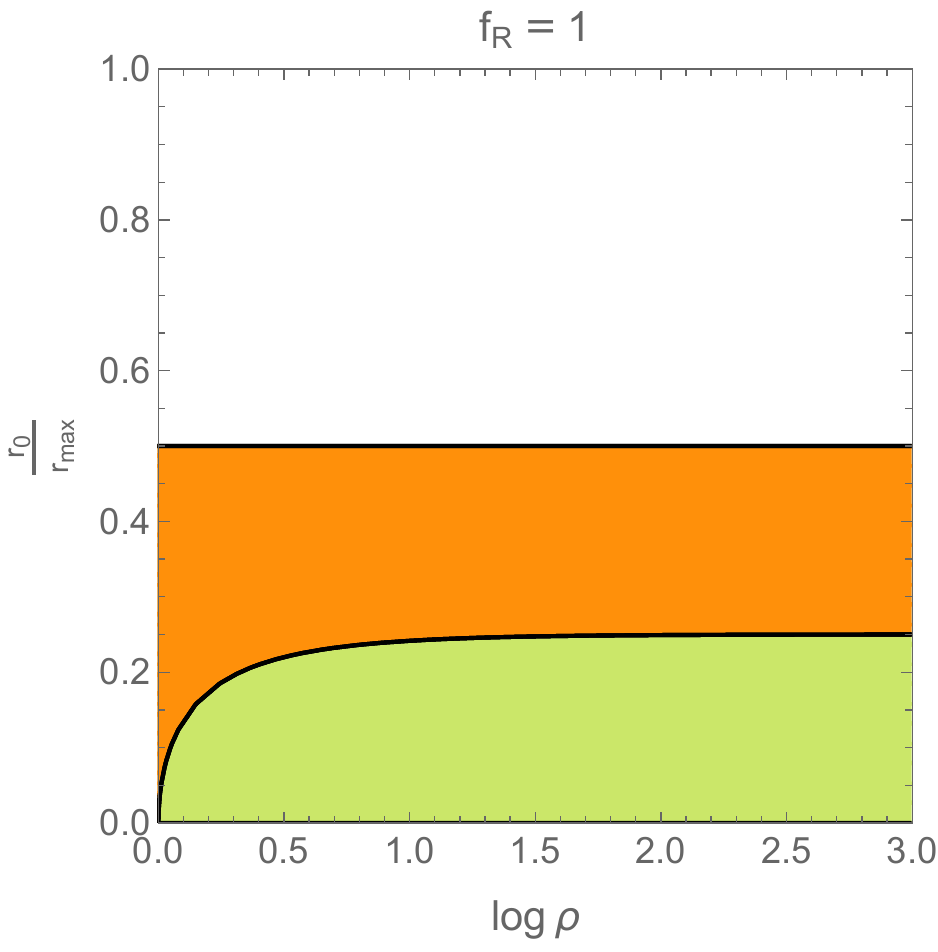}
    \caption{Behaviour of $r_{out}/r_{max}$ (green-orange boundary), $r_{pos}/r_{max}$ (orange-blue boundary), and $r_{pos}/r_{max}$ (blue-white boundary) with varying $\log \rho$ for fixed $f_R = \rho_R/\rho$ in the $K>0$ case, where $r_{max} = \pi/\sqrt{K}$ is the sphere diameter. In the shaded regions, the Lorentzian solution has a positive mass black hole. The Euclidean solution is sensible (the domain wall does not self-intersect) in the orange and green regions. The cosmological bubble is initially outside the horizon in the green region.}
    \label{fig:rNorm_vs_rho}
\end{figure}

For the $K > 0$ case with fixed $\rho$ and $f_R = \rho_R/\rho$ we generally have $0 \le r_{out} \le r_{int} \le r_{pos} \le r_{max}$, where the cosmological bubble is outside the horizon for $r_0 < r_{out}$, the Euclidean continuation of the solution makes sense (the domain wall does not self-intersect) for $r_0 < r_{int}$, the Lorentzian solution has positive mass if $r_0 < r_{pos}$, and $r_{max} = \pi/\sqrt{K}$ is the maximum value of $r$ corresponding to the sphere diameter. In Figure \ref{fig:rNorm_vs_rho}, we show the behaviour of $r_{out}/r_{max}$, $r_{pos}/r_{max}$, and $r_{pos}/r_{max}$ as a function of $\rho$ for a few different fixed ratios $f_R = \rho_R/\rho$. We note that in all cases, both Euclidean and Lorentzian solutions are sensible if the cosmological bubble includes less that half of the spatial sphere (i.e. $r_0 < \pi/(2 \sqrt{K})$. 

\paragraph{Arbitrarily large bubbles with $K=0$}

It is possible to show analytically that flat cosmology bubbles always have sensible Euclidean asymptotics, i.e. for any values of $r_0, \rho_R, \rho_M$ satisfying $K=0$. As we explained above, if the bubble is initially outside the horizon, the Euclidean solution does not include the point $R=R_H$ and so it is always a sensible geometry. 

On the other hand, if the bubble is initially inside the horizon, the solution has good Euclidean asymptotics only if condition \eqref{eq:good_Euc_asymptotics} is satisfied. For $K=0$ this condition is equivalent to the requirement $f_1(r_0,\rho_R) - f_2(r_0,\rho_R) > 0$, where we define
\begin{align}
    f_1(r_0,\rho_R) &\equiv \frac{2 \pi R_H}{1 + 3 R_H^2} \\
    f_2(r_0,\rho_R) &\equiv \int_1^{\sqrt{\rho_R}r_0} da\; g(a,r_0,\rho_R) \\
    g(a,r_0,\rho_R) &\equiv - \frac{a \left(a-\sqrt{\rho_R} r_0\right)}{\sqrt{(a-1) \left(a^3+a^2+a+\rho_R\right)} \left[a^3 r_0^2+a+r_0 \left((\rho_R-1) r_0-2 \sqrt{\rho_R}\right)\right]}.
\end{align}

We can obtain an analytic upper bound $f_2^\text{ub}$ of $f_2$ as follows
\begin{align}
    f_2 \le f_2^\text{ub} &\equiv  -\int_1^{\sqrt{\rho_R}r_0} da\; 
    \frac{a \left(a-\sqrt{\rho_R} r_0\right)}{\sqrt{(a-1) a^3} \left((a-1) \left(3 r_0^2+1\right)+\left(\sqrt{\rho_R} r_0-1\right)^2\right)}
    \\
    &=
    \frac{2}{y} \left(\frac{(x+y-1) \tan ^{-1}\left(\sqrt{\frac{y-(x-1)^2}{(x-1) x}}\right)}{\sqrt{y-(x-1)^2}}-\sinh ^{-1}\left(\sqrt{x-1}\right)\right),
\end{align}
where $x\equiv r_0\sqrt{\rho_R}$ and $y\equiv 1+3r_0^2$. Comparing this to the analytic expression for $f_1$ it is straightforward to check that $f_1(r_0,\rho_R) - f_2(r_0,\rho_R) \ge f_1(r_0,\rho_R) - f_2^\text{ub}(r_0,\rho_R) > 0$ for all values of $r_0$ and $\rho_R$. In particular, for large bubbles, we have
\begin{align}
    \lim_{r_0\rightarrow \infty} \frac{f_1}{f_2^\text{ub}}
    =
    \frac{\pi  \sqrt{3-\rho_R}}{3 (1-\rho_R)^{1/3} \sin ^{-1}\left(\frac{\sqrt{3-\rho_R}}{\sqrt{3}}\right)},
\end{align}
which is greater than 1 for all values of $\rho_R$. This shows that  geometries containing arbitrarily large bubbles of flat cosmology (with arbitrary matter to radiation density) have sensible Euclidean analytic continuations. It is interesting that we apparently have a Euclidean construction even for the extremely large flat bubbles where an entropy puzzle arises; in this case it would appear that any possible encoding of the possible states of the cosmological spacetime into the CFT would have to be non-isometric.

\section{CFT construction}
\label{sec:construction}

In cases where we have a sensible Euclidean construction, we can look at the asymptotics of this solution to understand the properties of a potential CFT dual construction. 

For the Schwarzschild part of the geometry, the asymptotic solution behaves as 
\begin{equation}
ds^2 = {1 \over Z^2} \left( dZ^2 + d {\cal T}^2 +  d \Omega^2 \right) ,
\end{equation}
where we have defined $Z = 1/R$. From this metric in Fefferman-Graham form, we can see that the boundary metric is conformal to part of a cylinder 
\begin{equation}
\label{eq:cylinder}
ds^2  = d {\cal T}^2 +  d \Omega^2 \qquad \qquad {\cal T} \in [- {\cal T}_0, {\cal T}_0] \; .
\end{equation}
The part of the Euclidean solution corresponding to the FRW regions has two asymptotic regions. We can represent the geometry of each of these in Fefferman-Graham form as 
\begin{equation}
    ds^2 = {1 \over z^2} \left(dz^2 + z_0^2( dr^2 + R_K^2(r) d \Omega^2) \right)
\end{equation}
where we have taken $|\tau| = \ln(z_0/z)$ for $\tau \to \pm \infty$. The boundary geometry is then conformal to 
\begin{equation}
\label{eq:BoundaryCap}
    ds^2 = z_0^2( dr^2 + R_K^2(r) d \Omega^2)
\end{equation}
which has the topology of a ball. Choosing $z_0 = 1/R_K(r_0)$ makes the $S^2$ at the boundary of this ball of unit radius so that the two balls corresponding to the two asymptotic regions of the Euclidean FRW geometry can be glued in to the two boundaries of the cylinder (\ref{eq:cylinder}). Thus, the full boundary geometry has the topology of a sphere. By a conformal transformation, this can be mapped to an infinite cylinder.

\paragraph{Constructing matter cosmologies by operator insertion}

As a simple case, we can consider the situation where there is no radiation and the matter is made up of heavy particles travelling along geodesics specified by fixed $(r, \theta, \phi)$ in the FRW spacetime. In the Euclidean picture, these geodesics continue to spatial geodesics, also at fixed $(r, \theta, \phi)$ extending between the two asymptotic regions. These particles can be obtained via the insertion of heavy operators in the CFT, similar to the construction of shells of matter in \cite{Anous:2016kss,Balasubramanian:2022gmo}.

If we consider particles of mass $m \gg 1/\ell_{AdS}$ in the bulk, the total number of these particles is
\begin{equation}
    n = {3 \over 8 \pi G \ell_{AdS}^2}{\rho_M V_K(r_0) \over m}
\end{equation}
where $V_K(r_0)$ is the proper volume of the FRW bubble given in (\ref{eq:volume}). 
For the CFT construction, we thus insert $n$ scalar operators with dimension
\begin{equation}
    \Delta \approx m \ell_{AdS}
\end{equation}
uniformly on the boundary geometries (\ref{eq:BoundaryCap}) associated with two asymptotic regions of the Euclidean FRW spacetime. Of course, we need to pick specific locations for these operators, but we can also imagine some ensemble of such insertion locations, so that the resulting state would be a mixed state corresponding to an ensemble of cosmologies with the matter particles in some ensemble of locations. In order to achieve an approximately uniform matter distribution, we can take the number of operators to be large while still requiring the corresponding operators to be heavy. This gives:
\begin{equation}
1 \ll m \ell_{AdS} \ll \left({\ell_{AdS} \over \ell_P}\right)^2 \; , 
\end{equation}
assuming that $r_0$ is of order $\ell_{AdS}$.

\begin{figure}
    \centering
    \includegraphics[width=0.4\linewidth]{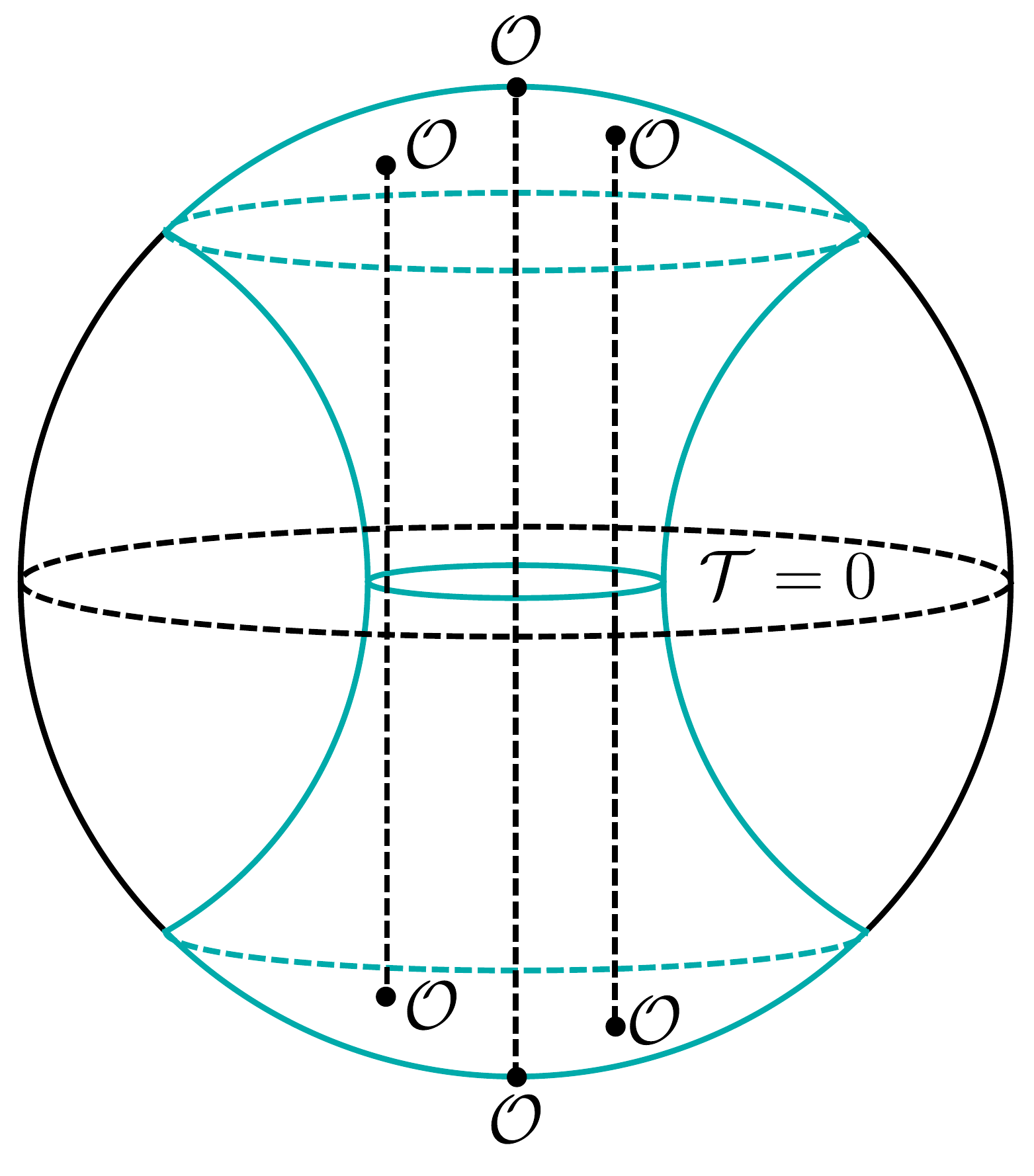}
    \caption{Preparation of the CFT state using correlated operator insertions in the Euclidean path integral. Operators are inserted in polar regions. Vertical dashed lines correspond to geodesics on which the matter particles live. Generally we may require an ensemble of such insertions.}
    \label{fig:EucInsrt}
\end{figure}

When considering a CFT path integral with operator insertions suggested by a certain Euclidean solution, it is important to understand whether the Euclidean solution of interest is actually the dominant saddle corresponding to that path integral. Otherwise, leading order calculations based on the CFT path integral might correspond to some other geometry that dominates the gravitational path integral. For the construction with the insertion of heavy operators, we are interested in a dual solution with Euclidean particle trajectories connecting the two ends of the wormhole. For these to be favoured over other solutions (e.g. where the geodesics connect operator insertions on the same side of the wormhole), it may be necessary to assume a correlated insertion of operators on the two sides (e.g. to consider a sum $\sum_i p_i {\cal O}_i(x_+) {\cal O}_i(x_-)$) where $x_+$ and $x_-$ correspond to the two ends of a geodesic, rather than just a single insertion ${\cal O}(x_+) {\cal O}_i(x_-)$ (see Figure \ref{fig:EucInsrt} for a schematic picture).
We plan to discuss this in more detail in a forthcoming publication.

\section{Probing the cosmology from CFT physics}

Given a CFT state describing a bubble-of-cosmology spacetime, it is interesting to ask how the spacetime can be reconstructed, e.g. to extract cosmological observables. 

A first observation is that in all of our spacetimes, the full spatial boundary is homologous to a point, so the entanglement wedge of the boundary is the full spacetime geometry. This suggests that the cosmological physics should be reconstructible in principle from the CFT  \cite{dong_reconstruction_2016, penington_entanglement_2019}.

On the other hand, reconstruction of black hole interiors in AdS/CFT is famously challenging, and in many cases, most or all of the cosmological region lies behind the black hole horizon. When the states we describe have a Euclidean CFT construction, this should not be a problem. Working in the Euclidean picture, it should be possible to use standard bulk reconstruction techniques (e.g. HKLL \cite{hamilton_local_2006,hamilton_holographic_2006}) to calculate observables in the Euclidean geometry. The Lorentzian observables are related to these by analytic continuation. 

Now suppose that we don't have a Euclidean construction or we wish to probe the cosmological physics directly from the Lorentzian state. Here, the recently proposed ``Python's Lunch'' criterion gives a precise suggestion for when the reconstruction will be difficult. According to the conjecture in \cite{Brown:2019rox} and  \cite{Engelhardt:2021mue}, the reconstruction of local bulk physics from the Lorentzian CFT state is highly complex when the region of interest lies in a Python's Lunch, that is, behind a non-minimal extremal surface. Our geometries have such a surface precisely when the bubble of cosmology lies completely behind the black hole horizon. In this case, the horizon at the $t=0$ slice is the non-minimal extremal surface. Spatially, this is a place where the $S^2$ areas begin to increase in size again as we move inward from the boundary. From our analysis above, the geometry will contain a python's lunch if and only if
\[
\rho_M + 2 \rho_R > 1 \qquad {\rm and} \qquad r_0 > r_c 
\]
where $r_c$ is the smallest value of $r_0$ such that
\[
R_K(r_0) = 1/\sqrt{2 \rho_R + \rho_M - 1} \; .
\]
In other cases, reconstruction of the entire spacetime should be simple; according to \cite{Engelhardt:2021mue} it can be accomplished by a combination of HKLL, the application of various Lorentzian sources and forward and backward evolution with the CFT Hamiltonian.

\paragraph{Probing with Ryu-Takayanagi surfaces}

In order to understand further how the cosmological spacetime is encoded in the boundary CFT state, it is useful to understand the Ryu-Takayanagi surfaces corresponding to various spatial subsystems of the boundary. These surfaces can indicate the extent of the entanglement wedge of a boundary spatial region of the CFT. The entanglement entropies of the boundary regions (suitably regularized e.g. by vacuum subtraction) also provide direct geometrical information about the interior via the usual Ryu-Takayanagi formula. 

\begin{figure}
    \centering
    \includegraphics[width=0.4\linewidth]{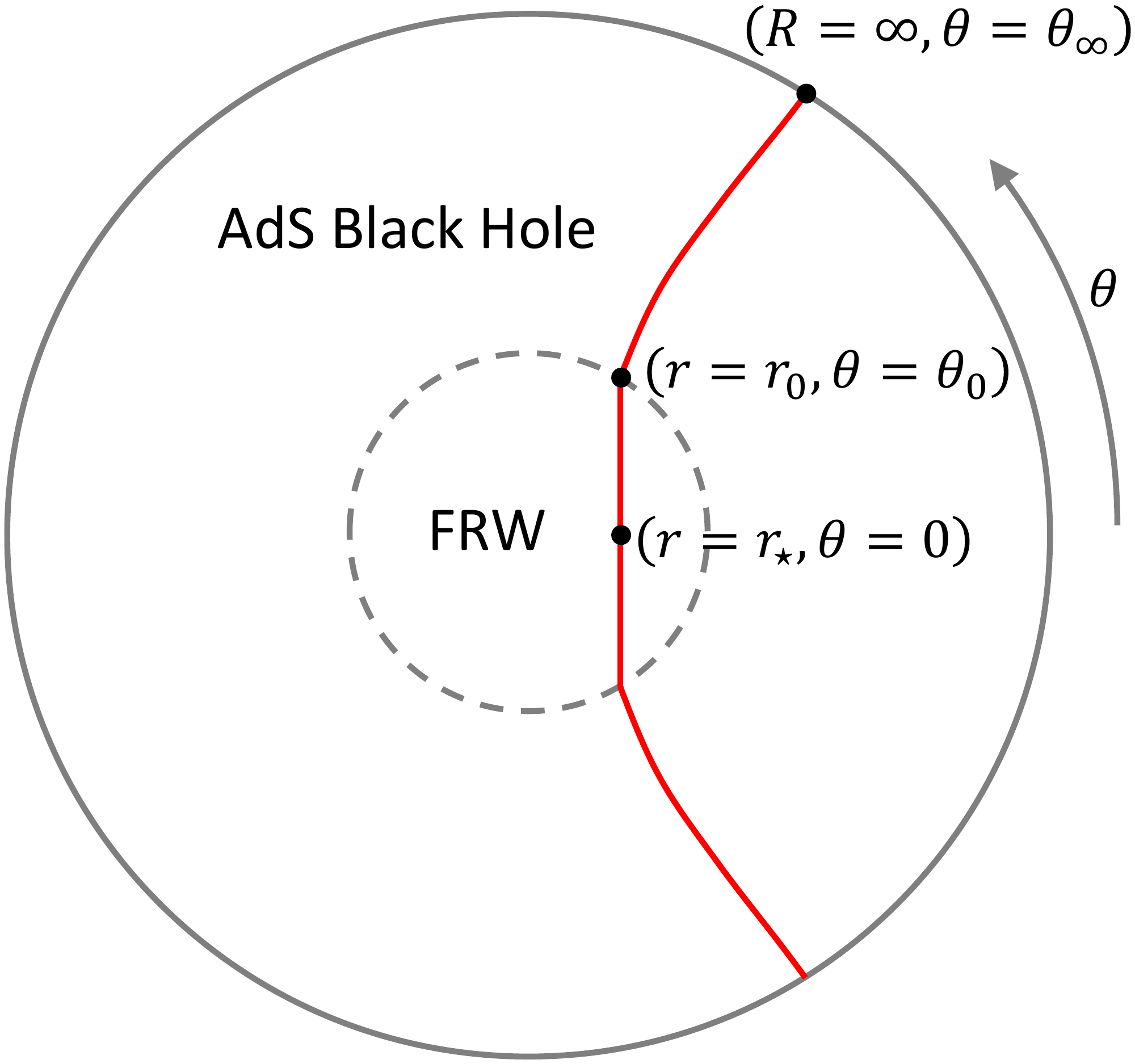}
    \caption{Schematic plot of an extremal surface at the time symmetric slice which probes the FRW region of the spacetime.}
    \label{fig:extremal_surface_schematic}
\end{figure}

We will focus on ball-shaped sub-regions $B$ of the $T=0$ spatial slice of the boundary spacetime. We can choose angular coordinates so that $\theta = 0$ is the center of the ball and the boundary of $B$ is $\theta = \theta_\infty$; see Figure \ref{fig:extremal_surface_schematic}. By the time-reversal and spherical symmetries of the bulk spacetime, the Ryu-Takayanagi surface is expected to lie on the time-symmetric slice in the bulk and preserve the rotational symmetry that leaves $B$ invariant. We can describe the RT surface trajectory as $R(\theta)$ in the Schwarzschild region and $r(\theta)$ in the cosmological region (if the surface extends into the bubble). In cases where the cosmology is completely inside the horizon, we need to consider separate Schwarzchild coordinate patches corresponding to the exterior region and the region past the horizon but outside the bubble. 

The RT surface is the surface homologous to $B$ that minimizes the area functional. The area functional may have multiple extremal surfaces homologous to $B$, in which case we need to compare the areas. It is convenient to parametrize the extremal surfaces by the location of their deepest point in the bulk (e.g. the distance $r_*$ from the center of the bubble), where $dr/d \theta = 0$ or $dR/d \theta = 0$. There will be a unique extremal surface for each such point, and we can integrate the extremal surface equations (below) to find the asymptotic behavior of this extremal surface. This gives some function $\theta_\infty(r_*)$, as schematically shown in Figure \ref{fig:extremal_surface_schematic}. 

When the surface lies entirely in the Schwarzschild exterior region, the area functional is 
\begin{align}
\nonumber    \Sigma(B) =  2\pi\!\int_{R_*}^{\infty} R \sin \theta \sqrt{R^2 d \theta^2 +\frac{d R^2}{F(R)}} \; ,
\end{align}
where we can take $\theta(R)$ or $R(\theta)$. When $r_*$ corresponds to a point in the FRW region, we have (assuming that the bubble is initially outside the horizon)
\begin{align}
\nonumber    \Sigma(B) = &2\pi\!\int_{r_*}^{r_0} dr R_K(r) \sin \theta_-(r) \sqrt{R_K^2(r) (\theta_-'(r))^2 + 1} \\
&+2\pi\!\int_{R_K(r_0)}^{\infty} dR R \sin \theta_+(R) \sqrt{R^2 (\theta_+'(R))^2 +\frac{1}{F(R)}} \;\notag .
\end{align}
where $\theta_-(r)$ and $\theta_+(R)$ parameterize the surface in the FRW and Schwarzschild regions respectively. Taking a variation $\theta \rightarrow \theta+\delta\theta$ of this functional with fixed $\theta_+(R= \infty) = \theta_\infty$ and $\theta_-(r_*) = 0$, we get
\begin{eqnarray}
\label{variation}
    \delta\Sigma &=& 2\pi \int_{r_*}^{r_0} \dd r \Big( R_K(r) \cos{\theta_-}\sqrt{L_-} - \frac{\dd}{\dd r}\frac{R_K^3(r)\sin{\theta_-}\theta_-'}{\sqrt{L_-}}\Big)\delta\theta \cr
    &&+ 
    2\pi \int_{r_0}^{\infty} \dd R \Big( R \cos{\theta_+}\sqrt{L_+} - \frac{\dd}{\dd R}\frac{R^3\sin{\theta_+}\theta_+'}{\sqrt{L_+}}\Big)\delta\theta_+ \cr
    &&+ \Big[\frac{R_K^3(r)\sin{\theta}\theta_-'\delta \theta_-}{\sqrt{L_-}}\Big]_{r_0} - \Big[\frac{R^3\sin{\theta_+}\theta'_+ \delta \theta_+}{\sqrt{L_+}}\Big]_{R_K(r_0)}  
\end{eqnarray}
where
\begin{equation}
    L_- = 
        R_K(r)^2 \theta_-'(r)^2+1,   \qquad
    L_+ = 
        R^2\theta_+'(R)^2+{1 \over F(R)}  \; .
\end{equation}
For the variation to vanish in general, the two integrands must vanish, giving differential equations for $\theta(r)$ and $\theta(R)$ in the FRW and Schwarzchild regions. Furthermore, the variations in the last line must also cancel, giving
\begin{equation}
\label{eq:interface}
    {\theta_-'(r_0) \over \sqrt{R_0^2 (\theta_-'(r_0))^2+1}} = {\theta_+'(R_0) \over \sqrt{R_0^2 (\theta_+'(R_0))^2+1/F(R_0)}} \; ,
\end{equation}
where we have defined $R_0 = R_K(r_0)$. This condition can be simplified to 
\begin{equation}
\label{eq:interfaceSimp}
    \theta_-'(r_0) = \theta_+'(R_0)\sqrt{F(R_0)}.
\end{equation}

In the FRW region, we have a symmetric space, so it is straightforward to derive an analytic expression for the extremal surfaces. For the surface with closest approach $r = r_*$ to $r=0$ at $\theta = 0$, we find\footnote{For positive curvature, we can describe the sphere as $\sum x_i^2 = 1/K$ in Euclidean space. An extremal surface with closest distance $r_*$ to $(1/\sqrt{K},\vec 0)$ is the plane $\vec{x} \cdot (-\sin(r_* \sqrt{K}),\cos(r_* \sqrt{K}),\vec{0}) = 0$. The change of variables $x_1 = \cos(r \sqrt{K})/\sqrt{K}, x_2 = \sin(r \sqrt{K})/\sqrt{K} \cos(\theta)$ maps back to our coordinates, giving the result. The flat and negatively curved results can be  obtained by taking $K$ to be zero or negative, respectively.}
\begin{equation}
\label{FRWRT}
\cos(\theta) = \left\{ \begin{array}{ll}   {\tan(r_* \sqrt{K}) \over \tan(r \sqrt{K})} & \qquad K > 0 \cr
   {r_* \over r}  & \qquad K = 0 \cr
  {\tanh(r_* \sqrt{|K|}) \over \tanh(r \sqrt{|K|})} & \qquad K < 0 
  \end{array}\right.
\end{equation}
To investigate the extremal surfaces starting at some radius $r_*$ in the FRW region, we can use the expressions (\ref{FRWRT}) to calculate $\theta_0 = \theta_-(r_0)$. The exterior portion of the RT surface can be determined by solving the differential equation for that region with $\theta_+(R_0) = \theta_0$ and $\theta_+'(R_0)$ obtained using (\ref{eq:interface}).
In cases where two different values of $r_*$ give the same $\theta_*$, we need to compare the areas of the surfaces and choose the one with smaller area.

\begin{figure}
    \centering
    \includegraphics[width=0.49\textwidth]{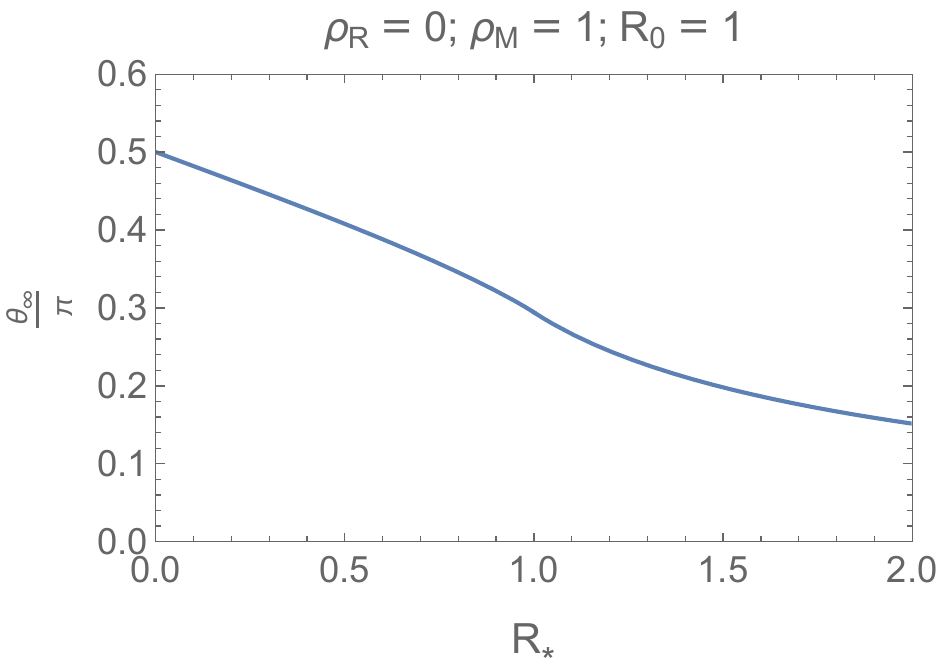}
    \includegraphics[width=0.49\textwidth]{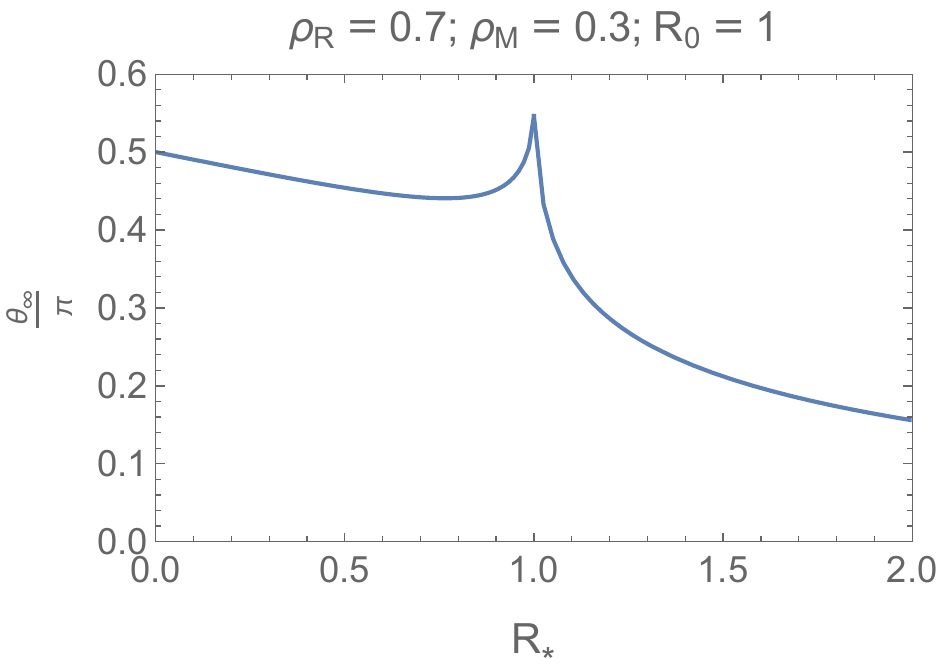}
    \includegraphics[width=0.49\textwidth]{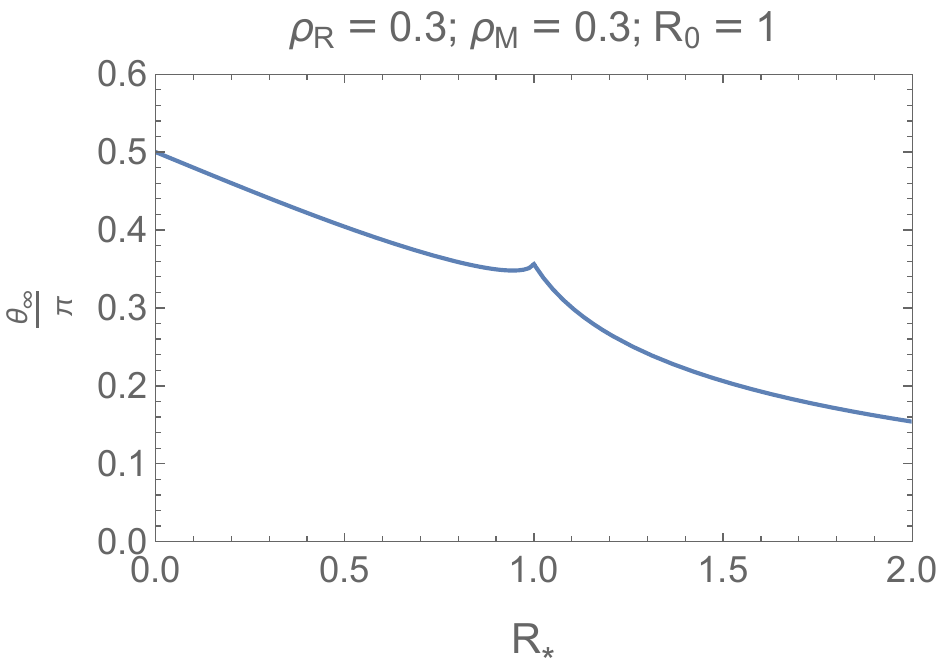}
    \includegraphics[width=0.49\textwidth]{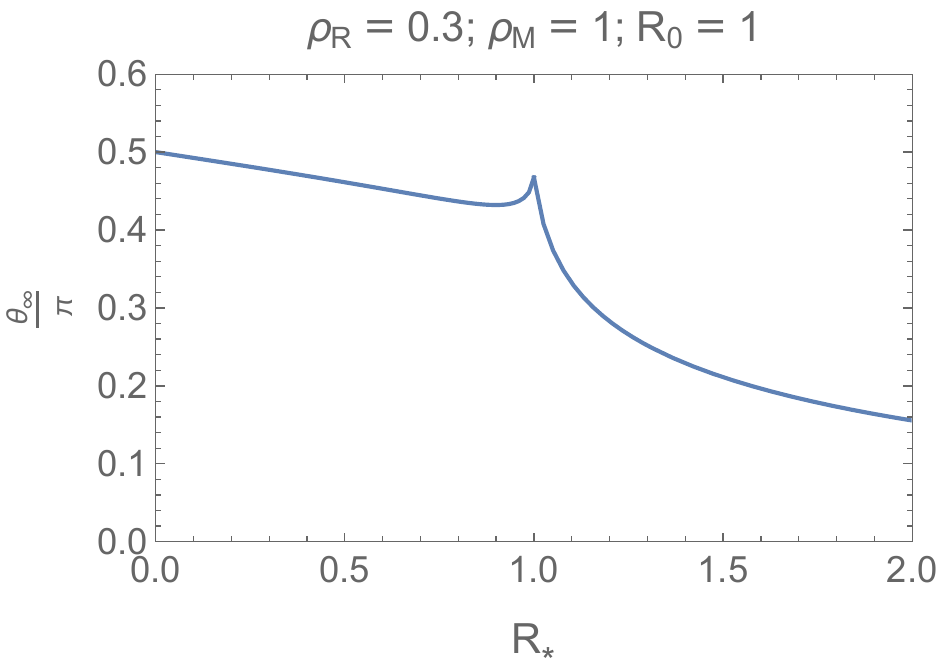}
    \caption{Angular size $\theta_\infty$ of spherical caps in the CFT, versus the smallest areal radius $R_*$ probed by corresponding extremal surfaces.}
    \label{fig:theta_vs_Rs}
\end{figure}

We show plots of $\theta_\infty$ as a function of $R_*$ in Figure~\ref{fig:theta_vs_Rs}, where we define $R_*$ to be the areal radius corresponding to the deepest point of the extremal surface. 
In these plots we fix the areal radius of the bubble separating the FRW region from the Schwarzschild region to be $R_0 = 1$. As a result $\theta_\infty(R_*)$ has a discontinuous first derivative at $R_* = 1$. Also notice that in all cases $R_* = 0$ implies $\theta_\infty = \pi/2$, i.e. the extremal surface passing through the centre of the FRW region corresponds to the hemispherical cap of the boundary CFT. 

Notice that for some parameter values (e.g. a flat bubble of dust; top left in the figure), $\theta_\infty$ is a monotonically decreasing function of $R_*$. In such cases the function $\theta_\infty(R_*)$ is invertible to $R_*(\theta_\infty)$, and thus there is a unique extremal surface homologous to a CFT cap of a given angular size. By the usual RT prescription this extremal surface must be the RT surface corresponding to the spherical cap. Furthermore, $R_*(\theta_\infty)$ ranges from $\infty$ to $0$ as $\theta_\infty$ ranges from $0$ to $\pi$, and hence the RT surfaces probe all parts of the spacetime, including all parts of the cosmological bubble.

More generically however (remaining plots in Figure~\ref{fig:theta_vs_Rs}), $\theta_\infty(R_*)$ fails to be monotonic for values of $R_*$ close to  $R_0$. This can occur for spatially flat FRW bubbles (top right), spatially hyperbolic bubbles (bottom left), and spatially spherical bubbles (bottom right). When there are multiple values of $R_*$ which give the same $\theta_\infty$, the areas of the surfaces must be compared and the RT surface would correspond to the minimum area extremal surface.

There are two possible manifestations of this non-monotonicity. 
In the first case when $\theta_{\infty}(R_0) < \pi/2$ there is some range $\theta_\infty \in (\theta_1,\theta_2)$, with $\theta_1,\theta_2<\pi/2$ where there are multiple candidate RT surfaces for the same $\theta_\infty$.
However, for large enough CFT balls, i.e $\theta_2<\theta_\infty<\pi/2$, there is a unique extremal surface (and hence RT surface) passing through the cosmological region. This is illustrated in the bottom two figures of Figure~\ref{fig:theta_vs_Rs}. 

In the second case the non-monotonicity is more severe, and $\theta_\infty(R_*)$ for $R_*>R_0$ covers the entire range $(0,\pi/2)$. An example of such a case is illustrated in the top right plot in Figure~\ref{fig:theta_vs_Rs}.  In this case surface areas of extremal surfaces would have to be compared in order to check whether any RT surfaces probe the cosmological bubble.



\section{Discussion}

In this paper, we have analyzed a family of exact solutions of $\Lambda < 0$ gravitational effective field theories where a comoving bubble of an FRW spacetime is embedded in an an exterior spacetime that is Schwarzschild-AdS. We found that for arbitrary matter and radiation densities and an arbitrary bubble size, the two parts of the spacetime can be patched across a thin shell of pressureless matter with density proportional to the square root of the interior radiation density. Physically, we can think of this as some perfectly reflecting dust. This construction appears to be special to 3+1 dimensions; in other dimensions, with both matter and radiation the required domain wall requires more complicated physics. Our choice to consider a domain wall with pressureless dust was purely for convenience; this allowed explicit analytic solutions that we could analyze. Of course, we could have more general solutions where the domain wall is not a thin shell. For example, there should exist a solution with only radiation, where the radiation is initially localized inside a ball-shaped region; while the radiation will spread out, the domain of dependence of the initial ball will still be FRW.

We found some promising signs that these cosmological bubble solutions, in some cases with arbitrarily large radius, can be dual to legitimate CFT states. The entropy of the exterior black hole is larger than the entropy associated with the cosmology except in the flat case for bubbles parametrically larger than the cosmological scale. The Euclidean continuations of the solutions appear to make sense for almost all cases and suggest a Euclidean CFT construction where operators and/or sources are inserted in ball shaped regions centered at opposite poles of an $S^3$ on which the CFT lives. We plan to investigate and discuss the Euclidean construction in more detail in a future work.

In various recent works with other collaborators, some of us have considered another approach to holographically describing $\Lambda < 0$ cosmologies. There, the idea was to holographically construct the Euclidean wormhole that is the analytic continuation of a full FRW cosmology. The connected wormhole solution with disconnected boundaries suggests that some type of ensemble or interaction between the CFTs should be involved. In \cite{VanRaamsdonk:2020tlr,VanRaamsdonk:2021qgv,Antonini2022}, we proposed an interaction between the CFTs involving a higher-dimensional field theory. In the setting of the present paper with a finite bubble of cosmology, the two asymptotic regions associated with the $\tau \to \pm \infty$ regions of the cosmological bubble are connected with each other via the asymptotic region of the Euclidean Schwarzschild geometry, so there is not a factorization puzzle. 
However, as we discussed in Section (\ref{sec:construction}) it may be that obtaining the desired saddle with an FRW bubble requires some correlation between the operator insertions on opposite sides of the sphere. In this case, taking the limit of an infinitely large bubble, we would end up with two disconnected CFTs, but with operator insertions correlated between the two CFTs. This would be a type of ensemble (over the operator insertions), but associated with a single CFT. The resulting picture is then similar to the construction in \cite{VanRaamsdonk:2020tlr,VanRaamsdonk:2021qgv,Antonini2022}, since there we could imagine integrating out the auxiliary theory, again leaving an ensemble of sources/operator insertions that are correlated between the two CFTs.

The cosmologies that we have considered are not realistic in that they have constant negative vacuum energy and never decelerate. However, as discussed recently in \cite{Antonini:2022fna}, more general solutions with a time-dependent scalar field could have an accelerating phase in the Lorentzian cosmology; such an accelerating phase produced by a time-dependent scalar appears to be consistent with the direct observations of scale factor evolution \cite{VanRaamsdonk:2023ion}. Constructing solutions with time-dependent scalars from the Euclidean picture would require additional sources for relevant scalar operators in the Euclidean CFT. These would need to have a particular dependence on the polar angle of the sphere for the scalar field homogeneous in the cosmological bubble. However, at this point, the main goal is to just able to study examples of four-dimensional big-bang cosmology using holography, so we are not particularly concerned with making these cosmologies realistic.

Having a fully microscopic description of cosmological spacetimes should allow in principle an investigation of questions that go beyond effective field theory. One such question is the nature of the big bang. In our construction, the big bang and big crunch singularities are special cases of past and future singularities in asymptotically AdS black holes.  A precise CFT description should also allow us to understand the quantum state of the cosmology and compute arbitrary cosmological observables. An interesting point is that the Euclidean constructions naturally construct a state at the time-symmetric point of the Lorentzian picture, rather than at the big-bang. Thus, for understanding the cosmological state away from the singularities, a detailed understanding of the big bang is not needed. Many of the conceptual issues associated with the holographic construction in this paper are shared with our other proposed construction in \cite{Antonini2022,Antonini2022short}. See those works for a more detailed discussion.

As a final comment, we note that for a given homogeneous and isotropic cosmology, the physics accessible to a single observer can likely be encoded holographically via dual CFT physics in many different ways, with various choices for the physics of the bubble wall and of the exterior spacetime. In \cite{simidzija_holo-ween_2020}, we considered bubbles of pure AdS spacetime (the trivial $\Lambda < 0$ cosmology), arguing that these could be described via many different possible dual CFTs by considering specific states of those theories defined by performing a certain Euclidean quench starting from the vacuum state of the CFT dual to the full AdS spacetime. Via a similar quench \cite{VanRaamsdonk:2018zws,van2020spacetime, may2021interpolating}, we can also encode the physics in the state of a collection of many non-interacting quantum systems. Similar constructions should exist for the more general cosmological spacetimes considered here. Thus, there may be many possible ways to holographically encode the physics accessible to a single observer in a cosmological spacetime. It would be interesting to understand if there is some more minimal description that avoids the excess baggage of an asymptotically AdS exterior spacetime. 

\section*{Acknowledgements}

We would like to thank Stefano Antonini, Brian Swingle, and Chris Waddell for helpful discussions.
We acknowledge support from the National Science and Engineering Research Council of Canada
(NSERC) and the Simons foundation via a Simons Investigator Award and the “It From
Qubit” collaboration grant.

\appendix

\bibliographystyle{jhep}
\bibliography{references}

\end{document}